\documentclass{article}
\usepackage[utf8]{inputenc}
\usepackage[numbers]{natbib}
\usepackage{amsmath,amssymb}

\usepackage[usenames,dvipsnames]{xcolor}

\usepackage{tcolorbox}
\usepackage{longtable}
\usepackage{booktabs}
\tcbuselibrary{skins,breakable}
\tcbset{enhanced jigsaw}

\definecolor{DarkRed}{rgb}{0.5,0.1,0.1}
\definecolor{DarkBlue}{rgb}{0.1,0.1,0.5}

\colorlet{YellowOrange}{RawSienna}

\usepackage{hyperref}
\hypersetup{
colorlinks=true,
pdfnewwindow=true,
citecolor=ForestGreen,
linkcolor=DarkRed,
filecolor=DarkRed,
urlcolor=DarkBlue
}
\usepackage{cleveref}

\usepackage{amsthm}
\newtheorem{lemma}{Lemma}
\newtheorem{claim}{Claim}
\newtheorem*{note}{Note}
\newtheorem{definition}{Definition}
\newtheorem{corollary}{Corollary}

\newtheorem{observation}{Observation}

\newtheorem{theorem}{Theorem}

\usepackage{graphicx}

\usepackage{xcolor}

\usepackage[flushleft]{threeparttable}

\usepackage{algorithmic}
\usepackage{algorithm}
\usepackage{comment}

\usepackage[shortlabels]{enumitem}

\newcommand{\E}{\mathbb{E}}

\newcommand{\cA}{\mathcal{A}}

\newcommand{\transcript}{\mathcal{T}}
\newcommand{\pol}{\mathcal{P}}
\newcommand{\lpopt}{OPT_{LP}}
\newcommand{\al}{\widehat{\alpha}}

\usepackage{mathtools}

\usepackage{tabularx}

\newcommand{\maxx}{\textsf{max}}
\newcommand{\minn}{\textsf{min}}

\newcommand{\gpa}{\text{Game Playing Algorithm}}

\newcommand{\Gpa}{\text{GPA}}

\usepackage{authblk}
\usepackage{fullpage}

\title{Efficient Stackelberg Strategies for Finitely Repeated Games}

\author{Natalie Collina}
\author{Eshwar Ram Arunachaleswaran}
\author{Michael Kearns}

\affil{University of Pennsylvania}

\affil[ ]{\{eshwar, ncollina, mkearns\}@cis.upenn.edu}

\date{February 2024}

\begin{document}

\maketitle
\begin{abstract}

We study Stackelberg equilibria in finitely repeated games, where the leader commits to a strategy that picks actions in each round and can be adaptive to the history of play (i.e. they commit to an algorithm). In particular, we study static repeated games with no discounting. We give efficient algorithms for finding approximate Stackelberg equilibria in this setting, along with rates of convergence depending on the time horizon $T$. In many cases, these algorithms allow the leader to do much better on average than they can in the single-round Stackelberg. We give two algorithms, one computing strategies with an optimal $\frac{1}{T}$ rate at the expense of an exponential dependence on the number of actions, and another (randomized) approach computing strategies with no dependence on the number of actions but a worse dependence on $T$ of $\frac{1}{T^{0.25}}$. Both algorithms build upon a linear program to produce simple automata leader strategies and induce corresponding automata best-responses for the follower. We complement these results by showing that approximating the Stackelberg value in three-player finite-horizon repeated games is a computationally hard problem via a reduction from balanced vertex cover.
\end{abstract}
\pagebreak 

\tableofcontents

\pagebreak 
\section{Introduction}

A Stackelberg game~\cite{Stack35} is a game in which one of the players, called the Leader, commits to a strategy which is communicated to the other players, called the Followers. The Stackelberg Equilibrium (SE) is the optimal strategy for the Leader to commit to, assuming the Followers best respond. This problem was first studied through a computational lens by Conitzer and Sandholm~\cite{conitzer2006computing}, who showed efficient algorithms in the case of two players and hardness results for three player games, and has since been the subject of many other works, including for settings such as security games~\cite{an2017stackelberg}. 

We consider Stackelberg equilibria in \emph{repeated}
 games, in which the Leader commits to an algorithm for playing the game at the start, and Followers best respond in algorithm space. Repeated games are a natural generalization of single-shot, simultaneous games --- agents repeatedly play the same game over multiple rounds, where the outcome in each round depends only upon the actions played in that round. While the Leader must commit to their algorithm upfront, this algorithm can be responsive to actions played in previous rounds. 

Much of the previous work in equilibria computation for repeated games has been in the infinite-horizon setting. Littman and Stone~\cite{littman2005polynomial} give an efficient algorithm for computing finite automata Nash equilibrium strategies achieving any given feasible payoff profile in infinitely repeated games. In the more recent work of Zhou and Tang~\cite{Zuo2015OptimalMS}, an analogous result was given for infinitely repeated Stackleberg games, again providing an efficient algorithm for computing finite automata strategies (and a generalization to automata with infinite states) for both the Leader and Follower. 

By contrast, in our work we focus on finite horizons. Reasoning about equilibria in finite-horizon games introduces many complexities not present in the infinite-horizon case. In infinite-horizon equilibria, any finite prefix of play can be ignored, and players can have threats of unbounded length. This is not true in finite-horizon games, and furthermore finite-horizon equilibria are sensitive to order of play and ``final-round'' complexities. The most relevant work in the space of finitely repeated Stackelberg games is~\cite{benoit1984finitely}. The authors provide a folk theorem for Subgame-Perfect Nash equilibria (SPNE) for finitely repeated games which has an implicit connection to SE computation, as we discuss in Section~\ref{sec:related}. However, the approach for finding Stackelberg equilibria implied by~\cite{benoit1984finitely} excludes a large class of games, and furthermore does not provide convergence rates. 


In our work, we provide the first algorithm that computes SE for \emph{all} finitely repeated games, including games with nontrivial SE which do not fulfill the specifications of~\cite{benoit1984finitely}, such as Prisoner's Dilemma. In addition, our algorithm constructing commitments has an approximation rate for every fixed $T$. Like previous papers on repeated games, the players in our game are allowed to employ a broad class of algorithms (essentially any finite time algorithm) to decide their strategies in each round based upon this history of play and their internal state. This allows for a large space of commitment schemes; for example, the Leader can employ ``threats" such as the grim-trigger strategy in the repeated Prisoner's Dilemma. We summarize our main results below:


\begin{itemize}[leftmargin=*]
    \item We give two efficient algorithms that find additively approximate Stackelberg strategies for the Leader that trade off different approximation factors. First, we show a $\left( \frac{2^{\text{poly}(n)}}{T}\right)$- additive approximate Stackelberg strategy where $n$ is the number of actions. 
    This is constructed using an algorithm that runs in polynomial time in the game size and $\log T$. We also present an approach using randomized sampling which gives a $O(\frac{1}{T^{0.25}})$-additive approximation that runs in time polynomial in the game size and $T$~\footnote{The approximation factors also depend upon the maximum number of bits needed to represent a single payoff.}. This second approach provides a slightly weaker approximation factor in $T$ in exchange for entirely avoiding dependence on $n$. Both constructions use simple automata (including states that play distributions over actions) for the Leader and also for a corresponding Follower best-response. We leave it as an open question whether there is an algorithm combining the best of both rates. 
    \item Building upon a reduction of Conitzer and Sandholm~\cite{conitzer2006computing} that showed that finding the Stackleberg value of a three player game is NP-hard, we show that even approximating the Stackelberg value of a three player repeated game upto an additive factor of $\frac{1}{T^{1/k}}$ is NP-hard (Theorem~\ref{thm:3_player_hardness}) for any integer $k$, ruling out any extension of our results for two player repeated games. We also use the same reduction to observe that approximating the Stackelberg value of three-player infinitely repeated games is NP-hard. 
    These results can be found in Section~\ref{sec:hardness_three}.
\end{itemize}

We also show a sequence of auxiliary results, that we discuss briefly here to help the reader to navigate through the paper.

\begin{itemize}[leftmargin=*] 
    \item We show that finding a Follower's best-response even to a poly-time Leader strategy is in general a computationally hard problem (Lemma~\ref{lemma:br_hard}). 
    However, our approximate Stackelberg Leader algorithm (Theorem~\ref{thm:stack_policy}) has an efficiently computable best-response (Lemma~\ref{lemma:br_poly}). 
    We discuss these results in Appendix~\ref{app:br_hard}.
    \item We show that the optimal Stackelberg Leader strategy may be different from simply playing the single-shot game's Stackelberg strategy in each round of play, through the example of the well known Prisoner's Dilemma (Section~\ref{sec:separation}).
    \item No-Regret algorithms are widely studied in the context of repeated games, due to their strong performance and strategic properties (See~\cite{blum2007learning},~\cite{freund1999adaptive}). A natural question is whether no-regret algorithms are in fact optimal Leader algorithms in the Stackelberg setting. We show that, somewhat unsurprisingly, arbitrary no-regret algorithms are in fact Stackelberg Leader algorithms in two player zero-sum games in Lemma~\ref{lemma:zs_nr_st}). Further, we also show that any no-regret algorithm is in fact a near best response to any other no-regret algorithms in repeated two player zero-sum games. 
    In contrast, we show via a counterexample that these properties do not hold when the zero-sum condition is removed, necessitating the broadening of the search to a more general class of algorithms. These results can be found in Appendix~\ref{app:nr}.
    \item Further, we generalize this result by showing that the class of learning algorithms (i.e. those algorithms that do not assume initial knowledge of the matrix, but learn the payoffs of each action after each round of play) does not always contain the optimal Stackelberg strategy for the Leader. In Appendix~\ref{app:learning_gap}, we show a simple pricing game where learning algorithms are strictly sub-optimal for the Leader to commit to.
\end{itemize}

\paragraph{\bf Our Techniques}
In proving our main result (Theorem~\ref{thm:stack_policy}) about efficiently computing a near optimal Stackelberg Leader algorithm, we use a sequence of ideas. Our main workhorse is a linear program (similar in spirit to that of~\cite{Zuo2015OptimalMS}) that is used to upper bound the average (per round) payoff that the Leader can guarantee themselves against a rational Follower. This linear program additionally gives us an ``ideal" empirical distribution over the action pairs that the transcript of play must follow to achieve the aforementioned value. We then convert this ideal empirical distribution into a concrete transcript of play for both the Leader and Follower. Akin to~\cite{littman2005polynomial},~\cite{Zuo2015OptimalMS} and~\cite{benoit1984finitely} we use the notion of ``threats" to stabilize such a transcript so as to force the Follower to play along with it. Our construction carefully choreographs the order of action pairs in this transcript to ensure that the Follower's best response is to follow the prescribed transcript. 
    
One might ask why we explicitly prescribe these ordered pairs, instead of allowing the Leader to draw pairs from the LP distribution i.i.d. each round. The problem is that even if the Leader commits to drawing from this distribution and playing their side of the strategy, they have no way to signal to the Follower what they should play at any given round before play occurs. Our approach, in spirit, involves the Leader pre-committing to these signals at the start of the game. However, this means that the Follower can see the payoffs of the specific move pairs they will be asked to play each round (as opposed to an expected payoff over an i.i.d. draw), which necessitates the ordering of pairs by increasing Follower payoff.
The transcript we construct forces through the worst action pairs for the Follower first, and adds ``treats" at the end to ensure that the Follower is incentivized to follow the prescription at each round of the game. The approximation to the upper bound comes from ``fitting'' a probability distribution to a finite number of rounds $T$, and improves as $T$ grows larger. While the average values of the Leader and Follower in the transcript will approximate their value in the LP distribution for large $T$, the ordering means that the actual transcript will not look like a typical sequence of i.i.d. draws from the distribution.
     
While the above approach gives us a rate of convergence to optimality that depends, possibly exponentially badly, on the number of actions for each player in the game, we show in Appendix~\ref{sec:faster} how to eliminate the dependence upon the number of actions. The main idea involved is a randomized sampling scheme that allows us to simplify the ideal empirical distribution found by the linear program. This process can also be viewed as randomized rounding of the linear program's solution to a more succinctly representable solution. This rounding also introduces randomization in the Leader's algorithm, and converts the absolute guarantee of near-optimality into a high probability guarantee. 
    
In Theorem~\ref{thm:3_player_hardness}, where we show the hardness of approximation of computing the Stackelberg value of repeated three player games, we demonstrate a gap-producing reduction from the balanced Vertex Cover problem. This reduction builds upon~\cite{conitzer2006computing}, which showed that finding the Stackelberg value of three-player games is NP-hard also via reduction from balanced Vertex Cover. Their hard instance does not directly extend to repeated games because of the richness of new strategies available to the Leader(s) in the repeated setting. Most of the work in our reduction goes into proving a stronger result about the single-shot case, which we do in Lemma~\ref{lemma:player3_min}. We can then lift this stronger result to the game with $T$ repeated rounds and prove hardness of approximation in Theorem~\ref{thm:3_player_hardness}. We also observe that the same lemma can be used to show that finding the approximate Stackelberg value of three-player infinitely repeated games is NP-hard.

\section{Related Work}

\label{sec:related}

The notion of Stackelberg Equilibria (SE) was first introduced by von Stackelberg in~\cite{Stack35}, who observed the advantage of strategic commitment in Cournot's duopoly model. Some followup work in the next decades did focus on applying the concept to repeated games, but these works focused on proving necessary and sufficient conditions for SE existence in various settings, rather than on constructive algorithms. ~\cite{Simaan73} explores SE in repeated games where the payoff matrix is either static or dynamic. They consider games where players have have an infinite number of moves (a setting where even Nash Equilibria may not exist) and derive conditions for SE existence.~\cite{Breton1985SequentialSE} consider \emph{sequential} Stackelberg games, repeated games where the state of the game matrix is affected by the moves that the players play each round. This setting is distinct from our own in that the Leader makes a commitment each round instead of committing to an algorithm at the start of the game. 

~\cite{benoit1984finitely} study Subgame Perfect Nash Equilibria (SPEs) in finitely repeated games and show that for a subset of these games, any feasible, individually rational pair of payoffs can be approximated to arbitrary precision for large enough time horizons. Given any such payoff pair, they give a method for constructing a SPNE that converges to this payoff value. As we show in our work, the payoff pair of any SE is indeed an individually rational and feasible in the game matrix. Thus, for games which satisfy ~\cite{benoit1984finitely}'s specifications, one could optimize over the convex set of feasible and individually rational payoff pairs (for the payoff of the Leader). One could then use their method to construct a pair of algorithms according to this value pair. The Leader could then commit to their side of the SPNE, which will converge to the Stackelberg value. 

While this observation is a promising sign that SE computation in finitely repeated games may be tractable, the above approach leaves a few things open. First of all, it will only work for a subset of repeated games;~\cite{benoit1984finitely}'s construction requires that all players have an equilibrium with a value above their minimax value. This excludes a large class of games with interesting finitely-repeated SE, such as Prisoner's Dilemma\footnote{The unique SPNE for Prisoner's dilemma is both players defecting in every round of play, via backward induction. Interestingly, we show in Section~\ref{sec:separation} that the Stackelberg value of the finitely repeated Prisoner's Dilemma is strictly larger than the value from the unique SPNE}. Second,~\cite{benoit1984finitely}'s result focuses on characterizing the set of SPNE and does not provide any guarantee on the rate of convergence. In particular, they prove that there \emph{exists} a $T_{0}$ for which their construction will converge to the desired approximation, but they provide no upper bound on how large this $T_{0}$ needs to be. Thus, there are no guarantees for the approximation rate of their construction for any fixed horizon $T$, which makes it infeasible to reason about any particular finite-horizon game.

 By contrast, we show an algorithm computing an approximate SE for all games, and provide explicit convergence rates for every tiem horizon $T$. Our results search beyond just the set of SPNE, and we explicitly use the power of the Leader to commit to playing sub-optimally in later rounds, something which is not possible in a SPNE (See Section~\ref{sec:separation} for an example).

In~\cite{conitzer2006computing}, Conitzer and Sandholm gave a polynomial algorithm for computing exact Stackelberg mixed strategies in single-shot, two-player games, bringing new algorithmic interest to the study of computing SE. Their method involves solving $n$ linear programs, with the $i$th program representing the best payoff that the Leader can get given that they are incentivizing the Follower to play pure strategy $i$. Picking the highest payoffs over all of these LPs gives the Leader their Stackelberg strategy for a single-shot game. They go on to prove that computing an exact Stackelberg strategy in a three-player game is NP-hard, via a reduction from Vertex Cover. 
A closely related work is~\cite{von2010Leadership}, which proves various properties about single-shot SE games with convex strategy sets.
These works led to further research considering the complexity of computing SE in various settings. The most closely related followup work involves NP-hard results for stochastic and extensive-form games, which are superclasses of finite-horizon repeated games, and a poly-time algorithm for \emph{infinite}-horizon repeated games. These results do not imply our results; more discussion on this can be found in
Appendix~\ref{app:rw}.

One recent and closely related line of work investigates \emph{learning} Stackelberg strategies in static repeated games. In the learning setting, the Stackelberg Leader begins with no knowledge of the payoff matrix, and only gains information via seeing their own payoff (and the opponent's move) after each round. Most works in this space assume that the opponent is responding myopically each round~\cite{balcan15},~\cite{Janusz2012},~\cite{lauffer2022}, and thus the Leader's goal is to learn and begin to play their single-round Stackelberg strategy.~\cite{haghtalab2022learning} studies a non-myopic (but discounting) Follower and a learning Leader, and develops a Leader algorithm via a reduction to bandit optimization in the presence of myopic agents. While the Follower is non-myopic, the authors crucially utilize the fact that the Follower is discounting in order to reduce the problem to a myopic-Follower setting. As with previous papers in the learning setting, their algorithms use the single-round Stackelberg value as a benchmark. 

Perhaps the best learning analogue of our work,~\cite{arunachaleswaran2024paretooptimal}, examines optimal commitment algorithms for finitely repeated games in the learning setting against a non-myopic Follower with no discounting. This setting is exactly the same as our setting, except the Leader does not have access to the Follower's payoff matrix. Being a Stackelberg Leader in the learning setting is significantly less powerful than in the full-information setting\footnote{To emphasize this point, we show a simple repeated pricing game with a fully myopic Follower in which the Stackelberg value of the Leader in the full information setting is strictly larger than any value the Leader can obtain in the learning setting. This can be found in Appendix~\ref{app:learning_gap}.}.~\cite{arunachaleswaran2024paretooptimal} instead introduce the solution concept of pareto-optimality over the space of payoffs of the Follower, and show that no-swap-regret algorithms, a subclass of no-regret algorithms, are optimal in this sense. In our work, we assume that the Leader has access to the payoff matrix of the Follower, and we are therefore able to achieve better payoff guarantees for the Leader than that in any work in the learning setting discussed here.



There is also work which reasons directly about the repeated-game setting, but considers the substantially distinct case of infinite-horizon repeated games. Zuo and Tang give a poly-time algorithm for computing SE in the space of machine strategies, which are automata that may have infinite states~\cite{Zuo2015OptimalMS}, and generalize their results to probability distributions over subclasses of these machine strategies. Similar to our work,~\cite{Zuo2015OptimalMS} solve an LP which optimizes the Leader's payoff when constrained to offer the Follower better than their safety value (called a security level in their paper). However, their proof heavily relies upon the infinite nature of the game. They present a Folk Theorem for infinite-horizon repeated games using machine strategies and use this to construct optimal commitments in the form of state machines. These state machines may have an infinite number of states, and therefore they do not explicitly represent them but rather show the existence of a Turing Machine which can do so. They also show additional results that examine how limiting the memory level of the Leader and Follower's machine strategies can affect game outcomes. Their work can be seen as a Stackelberg analogue of Littman and Stone's paper,
~\cite{littman2005polynomial}. Both works examine infinite-horizon repeated games, and both utilize constructive versions of Folk Theorems in order to construct strategies (in the space of algorithms) that are in (the respective type of) equilibrium. 

Work in finding equilibria in infinite-horizon repeated games cannot be directly applied to finite-horizon repeated games, as the algorithms for these two settings are structurally incomparable. While algorithms for infinite-horizon games have no sense of game length or game ending, algorithms for finite-horizon games can be constructed with the game length in mind. An algorithm for a finite-horizon repeated game can therefore behave differently, for example, in the final move of a game, or operate entirely differently depending on the initial game length. 

From a hardness perspective, Conitzer and Sandholm~\cite{conitzer2006computing} showed that finding the Stackelberg value of a three player game is $\textsc{NP}$-hard. Conitzer and Letchford~\cite{Letchford2010} showed that finding the Stackelberg value of a three player extensive-form game is NP-Hard. We extend the former and generalize the latter result by proving hardness of approximation for finite-horizon repeated three-player games. Borgs et al.~\cite{borgs2010myth} showed that no efficient algorithms exist for finding Nash Equilibrium in three player infinite-horizon repeated games unless $\textsc{PPAD}$ is in $P$ when players use algorithms that depend only upon the history of play (and not upon internal state). Halpern et. al.~\cite{halpern2014truth} later showed that this hardness result can be surmounted if the players are allowed to use (probabilistic polynomial time) algorithms that depend both upon the history of play as well as internal state (by demonstrating an efficient algorithm that finds a $\varepsilon$-approximate equilibrium). Their model is  an infinite-round analogue of our model (and that of~\cite{Letchford2010} for EFGs), which also allows the player's algorithms to generate actions using randomness, history of play and internal state.



In the context of playing repeated games using algorithms, we also mention no-regret algorithms, which were developed for decision making in online environments with experts (See~\cite{cesa2006prediction}). Multiple papers (See~\cite{freund1999adaptive},~\cite{blum1998line}) observed that these algorithms can be used by self interested agents to play repeated games due to the guarantees associated with them. In particular, in two player zero-sum games, the transcript of play converges to a Nash Equilibrium if all players use a no-regret algorithm, and to coarse correlated equilibria in general games. For special classes of games, such as atomic routing games/ congestion games (\cite{blum2006routing},~\cite{krichene2014convergence}) these are particularly compelling algorithms to use, since they guarantee convergence to Nash Equilibria and stability. However, as we show in Appendix~\ref{app:nr}, no-regret algorithms are not in general good algorithms for a Leader to commit to in the Stackelberg setting.


We summarize the main results in the field of finding equilibria in single-shot games, finite-horizon repeated games, and infinite-horizon repeated games in~\autoref{table:summary}. Our new contributions are \textbf{\textcolor{blue}{bolded and in blue}}.

\begin{table*}[]
\caption{Summary of known results} 
\begin{threeparttable}
\centering%
\begin{tabular}{|p{0.24\linewidth} |p{0.24\linewidth} |p{0.24\linewidth} |p{0.24\linewidth} |}
\toprule
& Single-shot & Finite Horizon & Infinite Horizon  \\
\midrule
2-player NE & Hardness of $o(1)$ -approximation~\cite{rubinstein2016settling} & Poly-time algorithm for Convergence to every SPNE~\cite{benoit1984finitely}, Hardness of $o \left (\frac{1}{T} \right)$- approximation\tnote{1} & Poly-time algorithm for exact solution~\cite{littman2005polynomial} \\
\midrule
3-player NE & Hardness of $o(1)$-approximation~\cite{rubinstein2016settling} & Hardness of $o\left (\frac{1}{T} \right)$- approximation\tnote{1} & Hardness of exact solution~\cite{borgs2010myth}, hardness of $\frac{1}{\textit{poly}(n)}$-approximation with no internal states~\cite{borgs2010myth}
\\
\midrule
2-player SE & Poly-time algorithm for exact solution~\cite{conitzer2006computing} & \textbf{\textcolor{blue}{Poly-time $ \min \left( \left( \frac{2^{\text{poly}(n)}}{T}\right), \frac{1}{T^{0.25}} \right)$ -approximation algorithm}} (Thm~\ref{thm:stack_policy})  & Poly-time algorithm for exact solution~\cite{Zuo2015OptimalMS} \\
\midrule
3-player SE & Hardness of $o(1)$-    approximation~\cite{conitzer2006computing}\tnote{2} & \textbf{\textcolor{blue}{Hardness of $\left (\frac{1}{T^{\frac{1}{c}}} \right)$-approximation}} (Thm~\ref{thm:3_player_hardness}) & \textbf{\textcolor{blue}{Hardness of $ \left (\frac{1}{\textit{poly}(n)} \right)$-approximation}} (Observation~\ref{infinite_3_hardness})~\tnote{3}\\
\bottomrule
\end{tabular}
  \begin{tablenotes}
    \item[1] These results come from the fact that, in finite-horizon repeated games, any pair of NE strategies must involve always playing a single-shot NE on the final round. The single-shot hardness result can therefore be used to implies that getting a $o(\frac{1}{T})$ approximation over $T$ rounds is PPAD-hard.
    \item[2] While their paper does not explicitly claim such a result in a lemma, their proof for hardness of finding an exact SE is in fact a hardness result for finding any approximate SE.
    \item[3] We do not provide a complete proof of this result, as the model we developed for this paper only makes sense for the finite-horizon case, but we provide a proof sketch and intuition.
  \end{tablenotes}
  \end{threeparttable}
  \label{table:summary}
\end{table*}

\section{Notation and Preliminaries}
\label{sec:prelims}
\begin{definition}[Bimatrix Games]
A bimatrix game $G$ is defined by two $n \times n$ matrices $M_1$ and $M_2$ that respectively denote the payoffs for Player $1$ and Player $2$ for each combination of actions chosen by the two players. We assume, without loss of generality, that both players have the same number of actions, and that their respective action sets are both indexed by the set $\{1,2,\cdots n\}$ denoted as $[n]$. If Player $1$ plays action $i$ and Player $2$ plays action $j$, then they get payoffs $M_1(i,j)$ and $M_2(i,j)$. In the asymmetric Stackelberg setting, Player $1$, or the row player, is assumed to go first and is referred to as the Leader while Player $2$, or the column player, is called the Follower. The strategies for Players $1$ and $2$ are points $x,y$ in $\Delta^n: = \{x \in \mathbb{R}^n_+ : \sum_i x_i = 1\}$ denoting the weights they place on the actions in their respective action sets. A pure strategy $i \in [n]$ is represented by the standard basis vector $e_i$. The expected payoff of Player $1$ (respectively $2$) is $x^\intercal M_1 y$ (respectively $x^\intercal M_2 y$). We assume that each entry in the payoff matrix is in $[-1,1]$ and is additionally an integer multiple of $\frac{1}{A}$, where $A$ is a sufficiently large integer. $A$ can be seen as a measure of the granularity of the payoff values and will become relevant when defining our approximation guarantees. 
\end{definition}

\begin{definition}[Nash Equilibria in Single-Shot Games]
A pair of strategies $x,y \in \Delta^n$ form a Nash Equilibrium for the game $G$ if there is no incentive for either player to deviate unilaterally, i.e. for all pure strategies $e_i$ ($e_j$) of Player $1$ (2), we have  $x^\intercal M_1 y \ge e_i^\intercal M_1 y$ ( $x^\intercal M_2 y \ge x^\intercal M_2 e_j$).
\end{definition}

\begin{definition}[Stackelberg Equilibria in Single-Shot Games]
A strategy $x \in \Delta^n$ is said to be a Stackelberg strategy for the Leader if it maximizes $x^\intercal M_1 y$ where $y \in \Delta^n$ is a best response for the Follower to $x$. Note that it suffices for the Follower to play a pure strategy $e_j$ as a best response. The induced play $(x,y)$ is called the Stackelberg Equilibria (SE) of the game. For the purposes of this paper, we assume that the Follower selects a best possible best response for the Leader if there are multiple best response strategies, a common assumption in Stackelberg literature. SE with this sort of tiebreaking assumption are known as \emph{Strong Stackelberg equilibria}, but for simplicity we will simply refer to \emph{SE} throughout this paper. 
\end{definition}


\begin{definition}[Repeated Bimatrix Game]
A repeated bimatrix game is defined by a bimatrix game $G$ and a time horizon $T$, both of which are known to all players. Players will play the game $G$ for exactly $T$ rounds, and their payoffs will be the sum of their payoffs over all rounds. The payoff matrix will remain static throughout all rounds. However, players may play different mixed strategies on different rounds and dynamically update their strategies based upon previous rounds of play. We refer to the pairs of actions $(i_1,j_2);(i_2,j_2),\cdots (i_T,j_T)$ played in the $T$ rounds as the transcript of the game and use $\transcript$ to represent this transcript. The values of interest to us are the {\bf expected average} per-round payoffs of the two players where the expectation is over the random bits used by the players. If $T = \infty$, then we call this an \emph{infinite-horizon} game. For much of this paper, we will consider finite $T$, in which case we call this a \emph{finite-horizon} game.
\end{definition}
It is obvious that players in a finite-horizon repeated game have a much richer action space than in a single-shot game, since their actions can depend upon the outcomes of the previous rounds, and even upon their internal states (possibly influenced by the random coins flipped in previous rounds). Therefore, we need to formalize exactly what this action spaces consists of -- in our paper, the players are assumed to delegate their play to algorithms, possibly randomized, that make their choices for them. We offer a general definition below.

\begin{definition}[$\gpa$]

A Game Playing Algorithm or GPA for a finite-horizon repeated game ($G$, $T$) is defined as a randomized algorithm that, in each round, takes as input the previous round's action pair and outputs an action for the current round. We allow this algorithm to have memory; in particular, it remembers what it needs to of the previous history of play as well as the random bits used to come up with actions in previous rounds.~\footnote{The fact that the algorithm can store previously used random bits allows for correlations in the probabilistic decisions made across rounds.}. Implicitly, it computes a distribution in $\Delta^n$ over the $n$ actions to be played in the $t$-th round, and then uses the randomness to select and play a particular action.


We say that a $\gpa$ is deterministic if it does not use any random bits. It is worth noting that each deterministic $\Gpa$ can be rewritten as/ shown to be equivalent to a look-up table , implying that operationally, the set of deterministic $\Gpa$s is finite. 
\end{definition}

To give further intuition for the richness of $\Gpa$s, consider some arbitrary game where the Leader player has two moves, $i_{1}$ and $i_{2}$, and the Follower player has two moves, $j_{1}$ and $j_{2}$. We provide some mechanics that $\Gpa$s support, along with (informal) descriptions of example Leader $\Gpa$s exhibiting these mechanics:

\begin{itemize}[leftmargin=*]
    \item History-dependency: play $i_{1}$ as long as the Follower plays $j_{1}$; if the Follower ever plays $j_{2}$, play $i_{2}$ for all remaining rounds.
    \item Time-dependency: play $i_{2}$ on the final round, and $i_{1}$ on all other rounds.
    \item Randomness: flip a new coin each round; play $i_{1}$ if heads, and $i_{2}$ if tails.
    \item \emph{Correlated} randomness: flip a coin at the start of the game. If heads, in all $T$ rounds of the game, flip a new coin, and play $i_{1}$ if heads and $i_{2}$ if tails. Otherwise, in all $T$ rounds of the game, play $i_{1}$.
\end{itemize}

\begin{observation}
The set of $\Gpa$s can equivalently be seen to be probability distributions over deterministic look-up tables. However, we keep this definition, since it offers a framework for succinct representations for these algorithms (for example - no-regret algorithms).
\end{observation}

We are almost ready to define a notion of SE for repeated bimatrix games. However, it is not a priori obvious that there exists a well defined best-response $\gpa$ for the Follower given a fixed $\Gpa$ for the Leader. We show that a best-response $\Gpa$ is in fact well defined, and that, in fact, it can be found among the set of deterministic (though potentially adaptive) $\Gpa$s. This result is analogous to the result in single-shot games that there always exists a best response pure strategy, and in fact uses this connection to work backward from the last round of play. Another way of seeing this result is to observe that finite-horizon repeated games can be rewritten as a large normal form game, implying the result. 
The proof can be found in the Appendix~\ref{app:bre}.

\begin{lemma}
\label{lemma:bre}
For any Leader $\Gpa$ by Player 1, there exists a best response $\gpa$ by Player 2 within the set of deterministic lookup table $\Gpa$s, which is a finite set. Therefore, a best response is well-defined in the $\Gpa$ space.
\end{lemma}





It is even less apparent that an optimal Leader algorithm to commit to is well defined -- we provide a constructive proof of existence. 
\begin{theorem}
\label{thm:compact_existence}
There exists an optimal $\Gpa$ $\pol$ for the Leader such that no other $\Gpa$, when paired up with its corresponding Follower best response, gives the Leader a better payoff than $\pol$ does against the corresponding Follower best response.
\end{theorem}

The key idea behind this proof is to observe that finite-horizon repeated games can be rewritten as normal form games, albeit at the cost of an exponential blow-up in the number of actions. This is not a new idea;~\cite{Letchford2010} makes the same observation in the context of Extensive-form games, which generalize finite-horizon repeated games. For sake of completeness, we sketch out the proof of this result in Appendix~\ref{app:compact}. We are finally ready to introduce the central solution concept of our paper.

\begin{definition}[Approximate Stackelberg $\Gpa$]
Let $P_{\max}$ be the payoff that an optimal Leader $\Gpa$ obtains against a best-responding Follower. Then a Leader $\Gpa$ $\pol$ is a $c$-approximate Stackelberg $\Gpa$ for any $c > 0$ if it obtains payoff at least $P_{\max} + c$. Note that the approximation definition only allows slack in the Leader's payoff--we still require that the Follower is \emph{exactly} best-responding.
\end{definition}

The precise computational question that we answer affirmatively in this paper is thus as follows: given a two player game $G$ (parameterized by the number of actions $n$ and the granularity $A$), and a time horizon $T$, does there exist an efficient algorithm to find an approximate Stackelberg $\Gpa$ for the Leader? 

\subsection{Separation of Repeated SE from Single-Round SE}
\label{sec:separation}

Even in a single-shot game, the Stackelberg Leader can often expect value higher than that of their best Nash equilibrium. As being a Leader in a single-shot game is so powerful, a natural question to ask is if repeated rounds can really strengthen the Leader's hand. In particular, one might observe that simply committing to play the single-shot Stackelberg strategy in each round will guarantee the Leader the Stackelberg value of the single-shot game on average. We address this question by showing an example where a Leader in a finite-horizon repeated game can do strictly better on average than getting their single-shot Stackelberg value each round. For our example, we focus on the simple and well-studied Prisoner's Dilemma game $\mathcal{PD}$. We will set the row player to be the Leader and the column player to be the Follower. The first move for both players represents cooperating, while the second move represents defecting. 

\[ \begin{bmatrix}
(3,3) & (0,5) \\
(5,0) & (1, 1)
\end{bmatrix} \]

\begin{lemma}
\label{lemma:separation}
The game $\mathcal{PD}$ repeated for $T$ rounds with $T \ge 3$ exhibits a constant separation between the Stackelberg value and the Leader value obtained by playing the single-shot game's Stackelberg strategy in each round.
\end{lemma}

The intuition is as follows: in the single-round case, a Stackelberg Leader can do no better than get payoff $1$, as the Follower has a dominant strategy of defecting. However, in a multi-round game, the Leader can incentivize the Follower to cooperate a majority of the time by promising to occasionally cooperate themselves, and threatening to defect if the Follower does not play along. We include a full proof of the above lemma, along with further discussion, in Appendix~\ref{app:gap_proofs}. 
We also include an explicit construction of an approximate Stackelberg $\Gpa$ for $\mathcal{PD}$ using our techniques in Appendix~\ref{app:pd_construction}.

\section{Algorithms for Approximate Stackelberg Equilibrium}
\label{sec:main_alg}

We now state and prove the main result of our paper, which gives efficient algorithms for computing 
approximate Stackelberg $\Gpa$s. 


\begin{theorem}
\label{thm:stack_policy}
We give two different efficient algorithms to compute approximate Stackelberg $\Gpa$s for a given bimatrox game repeated for $T$ rounds: 
\begin{itemize}
    \item An algorithm with running time polynomial in $n, \ \log A, \ \log T$ that finds a  $\left (\frac{2^{\text{poly}(n)} \cdot \text{poly}(A)}{T} \right)$-approximate Stackelberg $\Gpa$.
     \item A randomized algorithm with running time polynomial in $n,\  \log A, \ T$ that finds a  $O\left(\frac{\sqrt{A}}{T^0.25}\right)$-approximate Stackelberg $\Gpa$ (with high probability).
\end{itemize}
\end{theorem}

We begin by describing an LP for any bimatrix game which upper bounds the value guaranteed by any Leader $\Gpa$ for any $T$. Then, we describe a way to construct a $\Gpa$ using the LP solution such that a best-response by the Follower results in a Leader payoff closely approximating the LP value. We observe that, while the setup of our LP is different than that in \cite{Zuo2015OptimalMS}, the optimal value is always the same (for a certain regime in their paper). The value of the LP solution in \cite{Zuo2015OptimalMS} is equal to the value of the optimal SE strategy in infinite-horizon repeated games. Thus, in finding an approximation to the LP upper bound here, we are not only approximating the best Stackelberg strategy for any finite $T$ but also the best Stackelberg straetgy for an infinite-horizon repeated game.

\subsection{An LP Upper Bound}
\label{sec:LP_UB}
We describe the construction of the linear program for any bimatrix game. First, we precompute the ``threat" value $V$ of the column player along with the associated strategy for the row player $x^*$, which we call the threat strategy. This is defined as the minimum payoff that the column player can receive by best responding to the row player's strategy in the static setting i.e. $V : = \min_{x \in \Delta^n} \max_j x^\intercal M_2 e_j $ (recall that $M_2$ is the payoff matrix of the Follower/ column player). Note that this is equal to the value of a hypothetical zero-sum game between the two players where the row player's payoff is changed to be the negative of the payoff of the column player, and can hence be computed in polynomial time, as can the associated strategy $x^*$ of the row player. The reason we call this the threat value is that the row player could find a strategy that ensures the column player can do no better than $V$ against this strategy in any given round of play.

Next, we compute the solution of a linear program whose variables are weights $\{\alpha_{i,j}\}_{i,j}$ attached to each action pair in $\{i,j\}^2$. Intuitively, the point of the linear program is to find a prescribed probability distribution over the action pairs that maximizes the payoff of the Leader subject to the Follower receiving at least their threat value. 

\begin{flalign*}
    \max \sum_{i,j} M_1(i,j)\alpha_{i,j} \qquad \qquad  \alpha \in \mathbb{R}^n &\\
    \text{ subject to } &\\
    \sum_{i,j} M_2(i,j) \alpha_{i,j} \ge V &\\
     \sum_{i,j} \alpha_{i,j} = 1 &\\
    \alpha_{i,j} \ge 0 \qquad \qquad \text{ for } i,j \in [n]^2 &
\end{flalign*}

We observe that the above LP is feasible. In particular, the distribution over action pairs induced by the threat strategy $x^*$ of the Leader and the best response of the Follower is by construction a feasible point. Additionally, since the objective describes the payoff generated for the Leader by some distribution over the action pairs, the LP is bounded above by the maximum payoff of the Leader in the game. Consequently, the LP has a well defined optimum solution, which we also call $\alpha$ with value $\lpopt$.

\begin{lemma}
\label{lemma:lp_upper_bd}
The optimum value $\lpopt$ of the LP upper bounds the expected per-round payoff of a Stackleberg $\Gpa$ for any $T$.
\end{lemma}

The key intuition for this proof is that, no matter how intricate $\Gpa$ behavior may be thanks to randomness or adaptability, every pair of Leader and Follower $\Gpa$s induces \emph{some} distribution over sequences of move pairs. This distribution, which is analogous to the frequencies we solve for in our LP, determines the expected payoff for both players. If the expected payoff for the Follower from this move pair distribution is less than their safety value, they are certainly not best responding; thus, for this distribution to be feasible, it must adhere to our LP constraint. Given this constraint, our LP maximizes the payoff for the Leader. Therefore, any induced distribution which corresponds to expected payoff for the Leader cannot get value above our LP. A full proof is given in Appendix~\ref{app:lp_upper_bd}. 

\subsection{Construction of a Stackelberg $\Gpa$ from the LP}
\label{sec:construct_from_LP}
A short description of our candidate Stackelberg $\Gpa$ is as follows --- the Leader prescribes a sequence $\cA = a_1, a_2, \cdots a_T$ of action pairs for both Leader and Follower to adhere to. In case the Follower ever deviates from this prescribed sequence, the Leader then plays the threat strategy $x^*$ for the remaining rounds. This sequence is built based upon the LP solution $\alpha$ and has two key properties --- first, that the empirical distribution over action pairs in this sequence approximates the distribution $\alpha$ and second, that at any point in the sequence, the Follower never gains more by deviating from the sequence than following it.

The LP solution gives us the optimal distribution over action pairs that the Follower would be incentivized to play. If the Leader and Follower had a shared source of randomness, they could draw from this joint distribution each round and preserve both optimality for the Leader and incentive compatibility for the Follower. However, in our setting, there is no shared source of randomness or any mechanism for the Leader to pre-signal to the Follower what to play before each round. Therefore in our $\Gpa$ the Leader will prescribe a $T$-round sequence inspired by the LP solution, which preserves Follower rationality while approximating optimality. We call this $\Gpa$ $\pol^*(\alpha)$.

To aid with explicitly describing this sequence, we introduce some notation. We know that there is a polynomial time algorithm for solving a linear program with rational coefficients that outputs a rational solution(~\cite{schrijver2003combinatorial}). All coefficients in our linear program are rational --- in addition, the threat value $V$ is also a rational number, since it is itself the optimum of a linear program with rational coefficients. Therefore, we can assume that each $\alpha_{i,j}$ can be written in canonical form as $\frac{p_{i,j}}{q_{i,j}}$ where $p_{i,j},q_{i,j} \in \mathbb{N}$ and $gcd(p_{i,j},q_{i,j}) = 1$. For the next step, we re-index the action pairs from $1$ to $n^2$ such that for any two action pair $k_1,k_2 \in [n^2]$ with $k_1 < k_2$, we have $M_2(k_1) \le M_2(k_2) $. We use this indexing for the action pairs henceforth. Let $N = LCM(q_1,q_2 \cdots q_{n^2})$ and let $T = c \cdot N + r$ where $c$ is a natural number and $r \in \{1,2,\cdots N-1,N\}$. We assume $T$ is large enough to ensure $c \ge 1$. 
Additionally, we refer to $N$ as the cycle length, for reasons that will be clear from the full description of the $\Gpa$ $\pol^*(\alpha)$.

For each action pair $k$, we calculate the number of times $n_k$ to prescribe playing this action pair in the first $c \cdot N$ rounds, $n_k := \alpha_k \cdot c N = c \frac{p_k}{q_k} N$. By definition, $N$ is divisible by $q_k$ and hence $n_k$ is a natural number. Further, note that $\sum_k n_k = \sum_k \alpha_k \cdot c N = (\sum_k \alpha_k) \cdot c N  = c N$. The prescribed sequence for the first $c N$ moves is to play action pair $1$ for the first $n_1$ rounds, action pair $2$ for the next $n_2$ rounds and so on until the end of $c N$ rounds. 
For all the remaining rounds, the $\Gpa$ prescribes the action pair that maximizes the Follower's payoff. 

The intuition behind this ordering is that the $\Gpa$ forces through the more `painful' action pairs for the Follower at the beginning while promising rewards for cooperation and threats for defection. As we have required $r \geq 1$, the final round will always allow the Follower to get their optimal possible payoff if they have obeyed the sequence up to this point, which resolves any ``last-round'' incentive concerns.

This completes the description of the candidate Stackelberg $\Gpa$'s prescribed sequence. The logic of the $\Gpa$ itself will be to play according to the prescribed sequence, as long as the Follower has played according the the prescribed sequence in all previous rounds. If the Follower has ever deviated from the sequence, instead play $x^{*}$, the threat strategy, for all remaining rounds. This logic is enforced by the functions the $\Gpa$ is comprised of. A detailed example construction, which includes explicitly constructing the functions for the $\Gpa$, is shown in Appendix~\ref{app:pd_construction}. 



The following lemma uses backward induction and the ordering of the action pairs in the prescribed transcript to show that it is in the Follower's best interest to follow the prescription. The proof of this lemma can be found in Appendix~\ref{app:follow_presc}.

\begin{lemma}
\label{lemma:follow_presc}
Following the prescribed sequence for all $T$ rounds of play is a best-response $\Gpa$ (for the Follower) to the candidate Leader $\Gpa$ described above.
\end{lemma}

Next, we show a lower bound on the payoff of the Leader if the Follower follows the prescribed transcript.


\begin{lemma}
\label{lemma:reach_lp}
If the Follower obeys the prescribed sequence, the resulting payoff for the Leader will get a $2 N /T $ approximation to the optimum value $\lpopt$ of the LP.
\end{lemma}

\begin{proof}
In the first $c N$ rounds, the empirical frequencies of the action pair in the resulting transcript (when the Follower obeys the prescribed sequence) exactly equals the probability distribution $\alpha$. Therefore, the average payoff of the Leader in the first $c N$ rounds equals $\sum_{i,j} M_1(i,j) \alpha_{i,j}$, which is the objective of the LP for the optimal solution. In the remaining rounds, the difference in the average payoff and the LP's optimum objective is at most $2$ --- therefore, the total payoff is at least $\lpopt . (c N + r) - 2r$. Thus, we have a $2 r / T \le 2 N / T$- approximation to the LP optimum, where the last inequality comes from $r \le N$.
\end{proof}

Putting together Lemmas~\ref{lemma:lp_upper_bd} and \ref{lemma:reach_lp}, our candidate $\Gpa$ is shown to guarantee a payoff to the Leader that is a $ \frac{2 N}{T}$-approximation to the value generated by any $\Gpa$ against a rational Follower~\footnote{The Follower may choose some other equally good best response $\Gpa$ -- however in the definition of Stackleberg equilbria/ $\Gpa$, the Follower always picks the best possible option from the Leader's perspective while breaking ties, so the Leader's candidate $\Gpa$ is in fact guaranteed this approximation factor}. More generally, our approximation factor can be rewritten as $\frac{f(n,b)}{T}$ where $f(n,b) = 2^{\text{poly}(n,b)} = 2^{\text{poly}(n)} \cdot \text{poly}(A)$ since the number of bits used to represent $N$ is polynomial in $n$ and $b$ (recall that $b$ is the number of bits used to represent each payoff value) 

Additionally, the algorithm to construct the $\Gpa$ solves a linear program that is polynomial in the size of the game, and only does counting operations on $T$, and is therefore a polynomial in $n,b, \log T$ algorithm~\footnote{Finding an implicit representation of $\Gpa$ can be done in time polynomial in $\log T$, by, say, using a for loop. However, explicitly writing the $\Gpa$ as $T$-functions, one for each round, would still take time linear in $T$}.

We complement this result by showing in Appendix~\ref{app:inev} that it is impossible to exactly achieve the LP upper bound. Specifically, we show that a $\frac{1}{T}$-approximation to the LP upper bound is inevitable, showing that we have reached the limits of the LP based approach with respect to the dependence upon the time horizon $T$.

\subsection{Randomization for Faster Convergence}
\label{sec:faster}
We complete the proof of the theorem by showing that the judicious use of randomization can guarantee convergence to optimality which is independent of the number of actions $n$. In our current candidate $\Gpa$, the rate of convergence to optimality depends exponentially upon $n$. 


To get better dependence upon $n$, we use randomized sampling --- replacing the probability distribution $\alpha$ by the empirical average of a sufficiently large number of samples allows us to use standard concentration inequalities to approximate the Leader and Follower's values with high probability (since both of these are linear functions of the probability distribution). While simply approximating the Leader's payoff in this way is sufficient, we must ensure that the Follower's payoff is no lower than in the LP distribution. Thus, we tweak the empirical distribution slightly to ensure this. Finally, we need to add an extra ``treat'' for the Follower at the final round in order to ensure that they are incentivized to behave at every time step. Thus, instead of drawing $T$ samples, we will draw $T-1$ samples to leave room for this treat in the $\Gpa$ construction. Using this approach, we are able to create a $\Gpa$ which has strong guarantees for any $T$, independent of $n$ (with high probability).\\


Let us define $\al$ as the empirical distribution over action pairs constructed by drawing $T'=T-1$ I.I.D samples $a_1, a_2 \cdots a_{T'}$ from $\alpha$.

The payoff of both the Leader and Follower are a linear combination over the $T'$ random variables corresponding to the draws. Thus, we use the Hoeffding bound and the the union bound to claim that the following two inequalities hold with high probability (at least $1- \frac{4}{e^{50}}$ ).

$$
\left |\lpopt -  \frac{1}{T'} \sum_k M_1(a_k) \right | \le \frac{10}{\sqrt{T'}}
$$
$$
   \left | V - \frac{1}{T'} \sum_k M_2(a_k) \right | \le \frac{10}{\sqrt{T'}}
$$

While the first inequality adds to the approximation to optimality, the second inequality is a more serious matter. This is because the threat offered by the Leader is only credible if the prescribed transcript offers something no worse than the threat value. In particular, if we already have $V' := \frac{1}{T'} \sum_k M_2(a_k) \ge V$, then we can directly use this empirical distribution (at least for the first $T-1$ rounds). However, the other case demands some modifications to $\al$. In this event, $\al$ is tweaked in the following way: starting with the lowest-value pair for the Follower in the distribution, we will replace move pairs with the optimal move pair for the Follower (which gives the Follower value $m$). We will stop when the average value for the Follower over all the move pairs is at least $V$. We refer to the modified version of $\al$ as $\al'$. 

This process will certainly terminate with the Follower getting the required average value, as in the worse case we could simply make $T'$ swaps, resulting in the Follower getting expected value $m > V$~\footnote{The case of $m = V$ can be dealt with as a trivial special case}. 
By construction , the Follower's average value in $\al'$ is at least $V$.

However, each swap could replace something where the Leader gets payoff of up to $1$ with something where the Leader gets payoff of as little as $-1$. Thus, each swap could contribute up to $2$ to the Leader's total additive loss. $T'$ swaps would imply an additive loss of $2T' = 2T - 2$. Therefore, in order to retain an approximation for the Leader payoff, we must bound the number of swaps we will make in this process. \\

Let $c:=m-V$. Note that $c$ is always positive. Now, we will consider two cases: where $c \geq \frac{\sqrt{10}}{T'^{0.25}\sqrt{A}}$, and where $c < \frac{\sqrt{10}}{T'^{0.25}\sqrt{A}}$. 

In the first case, we claim that if this gap is large enough, then we only need to do relatively fewer swaps since each swap has a large effect, leading to a suitably small loss for the Leader.

In the other case, where $c < \frac{\sqrt{10}}{T'^{0.25}\sqrt{A}}$, we argue that the Follower's threat value is so close to their maximum possible payoff that most of the probability mass of $\al$ is already on the maximum payoff action pair, upper bounding the number of swaps possible.

Based on this reasoning, the following claim bounds the loss in Leader payoff due to the swaps.
The proof of this claim can be found in Appendix~\ref{app:few_enough_swaps}. 

\begin{claim}
\label{claim:few_enough_swaps}
The difference in average-per-round Leader payoffs between $\al$ and $\al'$ is at most $\frac{\sqrt{10A}}{T'^{0.25}}$.
\end{claim}

Similar to $\pol^*(\alpha)$, we can now put together a $\Gpa$ along similar lines based upon $\al'$ instead. We will order $\al'$ in increasing order of payoff for the Follower and prescribe actions in this order from round $1$ to round $T-1$. A reasonable question to ask is why we picked $T-1$ pairs instead of directly picking $T$ pairs.  
As with the construction in section 4.2, we must ensure that Follower is not tempted to deviate at any time step, even if there are few moves left and deviation will garner a significant payoff on that round. We do this by adding in a $T$-th move pair, which is another maximum payoff move pair for the Follower (garnering them a payoff of $m$). As in the previous Leader $\Gpa$ described, the Leader will begin playing the threat strategy $x^{*}$ every round if the Follower ever deviates.

Careful analysis along the lines of the proof of  Lemma~\ref{lemma:follow_presc} shows that the a best-responding Follower will obey the prescribed actions. Thus, the average payoff of the Leader will be their average payoff over the move pairs included in the Leader $\Gpa$. The Leader's payoff for the $T$th added move pair is at least $-1 \geq \lpopt -2 \geq \lpopt -2\sqrt{A} \geq \lpopt - \sqrt{10A}$. 
Calculating the total average payoff allows us to lower bound it by $\lpopt - \frac{4\sqrt{10A}}{T^{0.25}}$.

Thus, this sampling based approach completes the proof of Theorem~\ref{thm:stack_policy}.

Not only are both of the $\Gpa$s described efficiently computable, they have efficiently computable best-responses by the Follower. A proof of this claim can be found in the Appendix, in Lemma~\ref{lemma:br_poly}. This is not to be taken for granted; in Lemmas~\ref{lemma:br_hard} and~\ref{lemma:value_hard}, we show that for a general Leader strategy, even if it is restricted to run in polynomial time, finding a poly-time best-response for the Follower, and finding the expected value of the resulting game for the Leader, is NP-hard. We prove this hardness by embedding a 3-coloring instance into the repeated two player game problem.

\section{Hardness of Computing An Approximate Stackelberg GPA in 3-Player Games}
\label{sec:hardness_three}

While we have thus far considered only two-player Stackelberg games, the notion of an $n$-player Stackelberg game is also well-defined. In an $n$-player Stackelberg game, there is a strict ordering among all players. The first player in order makes a commitment, which all players get to see. Then the second player in order makes a commitment, and so on until the final player. In a single-shot game, this commitment is a (possibly mixed) strategy, while in the repeated games setting this commitment is a GPA. The SE is achieved when the first player in order commits to something which maximizes their expected value, given that all the remaining players best-respond in turn. The added complexity with more than two players is that, when making commitments, players must now reason about not just how a single Follower might respond to them but how the Follower might try to affect players further down the line. When we discuss additive approximations, we are still referring only to the approximation to the payoff of the very first player in line. All Followers are assumed to be perfectly best-responding.


We must be careful when discussing hardness in algorithm space; we cannot simply say that constructing an optimal GPA is hard just because finding the optimal move commitments in a GPA is hard. This is because GPAs are composed of algorithms, so the hard computational work could be delegated to the GPA itself. Thus, our hardness result for three-player approximation tells us two things: 1) finding the approximate \emph{value} of a Stackleberg equilibrium in the three-player setting is hard, and 2) finding an approximately optimal \emph{GPA} for Player 1 in the three-player setting is hard, as long as all GPAs are restricted to run in poly-time.

The hardness result we build upon on is from \cite{conitzer2006computing}, where Conitzer and Sandholm prove that computing a single-shot SE for three-player games is NP-hard. While not explicitly mentioned in their top-level results, their proof in fact shows that approximating a single-shot three-player SE is also hard. This hardness of approximation result does not immediately imply hardness of approximation for the repeated setting; as we have previously shown in the two-player case, the Stackelberg GPA may look very different from the single-shot Stackelberg. Thus we cannot perform a trivial reduction from the single-shot problem to the repeated-game problem. However, we can still utilize Conitzer and Sandholm's work to our benefit by looking closely at their results, and proving that their hard instance in fact extends to approximation in repeated games. 

\begin{theorem}
\label{thm:3_player_hardness}
In three-player finite-horizon repeated games over $T$ rounds, there exists no polynomial in $n,A,T$ time algorithm computing a $\left (\frac{A^k}{T^{\frac{1}{k}}} \right)$-additive approximation~\footnote{An earlier version of this paper incorrectly claimed to rule out algorithms with a  $\Omega \left (\frac{A^k}{T^{\frac{1}{k}}} \right)$ additive approximation, the statement of the result is therefore corrected, although the proof, and implication, of the result remains unchanged } to the Stackelberg value of the game unless $P =NP$. Here, $k$ is any natural number.
\end{theorem}






Conitzer and Sandholm prove hardness for three-player single-shot games via a reduction from balanced vertex cover. They construct a three-player payoff matrix from a graph for which Player 1 can get a nonzero payoff iff there exists a vertex cover in the graph of size $\frac{|V|}{2}$. We present the input transformation they utilize in the box below:

\noindent\fbox{%
    \parbox{\textwidth}{%
        Function $f$\\
        Input: Graph $G = (V,E)$ with $n$ vertices and $m \le n^2$ edges\\
        Output: A three-player game matrix where Players 1 and 2 have $n$ pure strategies and Player 3 has $m + n + 1$ pure strategies. \\
        Construction: Players 1, 2 and 3 will all have a pure strategy associated with every vertex in the graph: $r_{v}$, $s_{v}$ and $t_{v}$ respectively. In addition, Player 3 will have a pure strategy associated with every edge in the graph, $t_{e}$. Finally, Player 3 will have one last strategy $t_{0}$. The utilities are listed below:\\
        $\mu_{1}(r,s,t_{0}) = \mu_{2}(r,s,t_{0}) = 1$\\
        $\mu_{1}(r,s,t_{k}) = \mu_{2}(r,s,t_{k}) = 0$, $k \neq 0$\\
        $\mu_{3}(r, s, t_{0}) = 1$\\
        $\mu_{3}(r_{v}, s_{v}, t_{v}) = \frac{n}{n-2}$, $t_{v} \neq r_{v}$ and $t_{v} \neq s_{v}$\\
        $\mu_{3}(r_{v}, s_{v}, t_{v}) = 0$, $t_{v} = r_{v}$ and/or $t_{v} = s_{v}$ \\
        $\mu_{3}(r_{v},s_{v},t_{e}) = \frac{n}{n-2}$, $r_{v}$ not adjacent to $t_{e}$ \\
        $\mu_{3}(r_{v},s_{v},t_{e}) = 0$, $r_{v}$ adjacent to $t_{e}$ 
    }%
}

Note that this input transformation works in time polynomial in $n$. Conitzer and Sandholm used this transformation to reduce from Balanced Vertex Cover to Exact One-Round Three-Player Stackleberg. We will use the same transformation $f$ in our reduction to reduce from BALANCED-VC, but our correctness proof will require some additional heavy lifting. We also additionally define $T$ to be some suitably chosen polynomial in $n$. 

Let us formally define the problem BALANCED-VC. The problem takes as input a graph $G$. It outputs YES if the graph has a vertex cover of size at most $\frac{n}{2}$. It outputs NO if the graph does not have any vertex covers of size at most $\frac{n}{2}$. This problem is known to be NP-complete, through a simple reduction from the Vertex Cover problem~\cite{karp1972reducibility}. 


Consider the decision problem APPROX-STACK, that takes as input a three player game $G$ with $n$ actions per player and playoffs of granularity $1/A$, a time horizon $T$ and a real number $V$. The problem is to distinguish between games that have Stackelberg value greater than or equal to $V$ (Yes) and games that have Stackelberg value less than or equal to $V - \frac{A^k}{T^{\frac{1}{k}}}$ (No).

We will now construct a reduction from BALANCED-VC to APPROX-STACK. 
The reduction will be as follows: on input graph $G$, apply the transformation $f$ described above to get a game matrix $M$, then define a problem in APPROX-STACK as $(M,T)$ with value $V=1$ and $T = n^{ck + k^2 }$ with $c = 5$. We note that the granularity $A$ of the game is $n-2$, upper bounded by $n$.
If the Stackelberg value is greater than or equal to $1$, we answer that there is a vertex cover of size $\frac{|V|}{2}$ and if the Stackelberg value is less than $1 - \frac{A^k}{T^{\frac{1}{k}}}$, we will answer that there is no vertex cover of size $\frac{|V|}{2}$. Note that $1 - \frac{A^k}{T^{\frac{1}{k}}} > 1- \frac{n^k}{n^{c + k}} = 1 - \frac{1}{n^{c}}$, and therefore such an algorithm for APPROX-stack will always answer that there is a balanced vertex cover if the Stackelberg value is $\geq 1$ and always answer that there is a no balanced vertex cover if the Stackelberg value is less than $1 - \frac{1}{n^{c}}$. 

We will now prove that this is a valid reduction via the following lemmas.

\begin{lemma}
\label{lemma:first_half_hardness}
If there exists a vertex cover of size $\frac{n}{2}$ in $G$, then the game $M,T$ has a value of at least 1. 
\end{lemma} 

The proof of this lemma, the easier half of our proof, builds directly upon the ideas in~\cite{conitzer2006computing}. The proof can be found in Appendix~\ref{app:first_half_hardness}.

Proving the other half is significantly harder. To aid in this proof, we first claim an auxiliary result, the proof of which appears later. This result shows that when every vertex cover is of size at least $n/2+1$, then Player 3 always has a response to any strategies employed by the first two players in a given round that guarantees a payoff that is inverse polynomially better than 1. 

\begin{lemma}
\label{lemma:player3_min}
If there exists no vertex cover in $G$ with size $\frac{n}{2}$, then for every pair of mixed strategies by Players 1 and 2 in a given round, Player 3 can always play a move that garners them an expected value strictly greater than $1 + \frac{1}{(n-2)(n^{c-1})}$
\end{lemma}

We use this critical observation to prove the below lemma and complete the proof of Theorem~\ref{thm:3_player_hardness}. The proof of this lemma can be found in Appendix~\ref{sec:player3_min}.

\begin{lemma}
If there does not exist a vertex cover in $G$ with size $\frac{n}{2}$, then the game $M,T$ has Stackelberg value strictly less than $1 - \frac{1}{n^{c}}$.
\end{lemma} 
\begin{proof}
Consider all of the $n$ moves that Player 1 can play, each associated with a vertex from the original graph $G$. 

Consider any triplet of $\Gpa$s that Players 1 and 2  commit to (in order) and player 3 best responds to. 
Note that Lemma~\ref{lemma:player3_min} implies that Player 3 must get expected value (averaged over the rounds) strictly greater than $1 + \frac{1}{(n-2)(n^{c-1})}$ in this equilibrium (since they could obtain this value by just picking the best response in each round myopically).

$$\E \left [\sum_{t \in T}\textit{Payoff of Player 3} \right] > 1 + \frac{1}{(n-2)n^{c-1}}$$
By linearity of expectation:
$$\sum_{ t \in T}\E \left[\textit{Payoff of Player 3} \right] > 1 + \frac{1}{(n-2)n^{c-1}}$$

Let Player $3$ play action $t_0$ with probability $p_t$ in round $t$. Then, $\frac{1}{T} \sum_t p_t$ is the fraction of time Player 3 plays $t_{0}$ and $1 -\frac{1}{T} \sum_t p_t$ is the expected fraction of time Player 3 does \emph{not} play $t_{0}$. In the case of Player 3 playing $t_{0}$, they will get a payoff of $1$. Otherwise, they will get a payoff of at most $\frac{n}{n-2}$. Thus, we can write

$$\E \left [\sum_{t \in T}\textit{Payoff of Player 3} \right] \leq 1 \cdot \frac{1}{T} \sum_t p_t + \frac{n}{n-2} \cdot \left(1 -\frac{1}{T} \sum_t p_t \right)$$

Putting our two inequalities together, we have

$$1 \cdot \frac{1}{T} \sum_t p_t + \frac{n}{n-2} \cdot \left(1 -\frac{1}{T} \sum_t p_t \right) > 1 + \frac{1}{(n-2)n^{c-1}}$$

Rearranging:
$$\left(\frac{2}{n-2} \right) \left(\frac{1}{T} \sum_t p_t \right ) < - 1 - \frac{1}{(n-2)n^{c-1}} + \frac{n}{n-2}$$
$$\frac{1}{T} \sum_t p_t < - \frac{n-2}{2} - \frac{1}{2n^{c-1}} + \frac{n}{2}$$
$$\frac{1}{T} \sum_t p_t < 1 - \frac{1}{2n^{c-1}} $$

Note that as Player 1's payoff every round is determined solely by whether or not Player 3 plays $t_{0}$, Player 1's payoff can be written as
$$  1 \cdot \frac{1}{T} \sum_t p_t $$

So, Player 1's expected average payoff (averaged over $T$ rounds) is strictly less than $1 - \frac{1}{2n^{c-1}}$, which for $c \geq 2$ and $n \geq 2$ is $\leq 1 - \frac{1}{n^{c}}$, completing our proof. 
\end{proof}

\begin{observation}
\label{infinite_3_hardness}
For infinite-horizon Stackeleberg games with three players, finding an approximate equilibrium is NP-hard. 
\end{observation}

While the mechanics of $\Gpa$s are explicitly tailored to deal with finite-horizon games, we observe that a similar line of reasoning implies hardness of approximation for the infinite-horizon case, as long as the strategies used by the players have well-defined limit-averaged payoffs. If there exists a pair of distributions by Players 1 and 2 that incentivize $t_{0}$ on any single round, they can commit to playing these distributions infinitely, ensuring that Player 1 gets an average payoff of $1$. If there does not exist such a pair of distributions, then by Lemma~\ref{lemma:player3_min}, Player 3's time-averaged payoff for playing $t_{0}$ every round will always be well-separated from some other move by some $\frac{1}{\text{poly}(n)}$. Therefore, they must play moves other than $t_{0}$ with frequency 
(in the limit average) $\frac{1}{\text{poly}(n)}$, which implies that Player 1's limit-average payoff will be bounded away from $1$ by $\frac{1}{\text{poly}(n)}$.


\bibliography{references}

\newcommand{\etalchar}[1]{$^{#1}$}
\begin{thebibliography}{HLNW22}

\bibitem[ACS24]{arunachaleswaran2024paretooptimal}
Eshwar~Ram Arunachaleswaran, Natalie Collina, and Jon Schneider.
\newblock Pareto-optimal algorithms for learning in games, 2024.

\bibitem[ATS17]{an2017stackelberg}
Bo~An, Milind Tambe, and Arunesh Sinha.
\newblock Stackelberg security games (ssg) basics and application overview.
\newblock {\em Improving Homeland Security Decisions}, page 485, 2017.

\bibitem[BAH85]{Breton1985SequentialSE}
Michel~Le Breton, A.~Alj, and Alain Haurie.
\newblock Sequential stackelberg equilibria in two-person games.
\newblock {\em Journal of Optimization Theory and Applications}, 59:71--97,
  1985.

\bibitem[BBH{\etalchar{+}}17]{Bosansky2017}
Branislav Bo\v{s}ansk\'{y}, Simina Br\^{a}nzei, Kristoffer~Arnsfelt Hansen,
  Troels~Bjerre Lund, and Peter~Bro Miltersen.
\newblock Computation of stackelberg equilibria of finite sequential games.
\newblock {\em ACM Trans. Econ. Comput.}, 5(4), dec 2017.

\bibitem[BBHP15]{balcan15}
Maria-Florina Balcan, Avrim Blum, Nika Haghtalab, and Ariel~D. Procaccia.
\newblock Commitment without regrets: Online learning in stackelberg security
  games.
\newblock In {\em Proceedings of the Sixteenth ACM Conference on Economics and
  Computation}, EC '15, page 61–78, New York, NY, USA, 2015. Association for
  Computing Machinery.

\bibitem[BCI{\etalchar{+}}10]{borgs2010myth}
Christian Borgs, Jennifer Chayes, Nicole Immorlica, Adam~Tauman Kalai, Vahab
  Mirrokni, and Christos Papadimitriou.
\newblock The myth of the folk theorem.
\newblock {\em Games and Economic Behavior}, 70(1):34--43, 2010.

\bibitem[BEDL06]{blum2006routing}
Avrim Blum, Eyal Even-Dar, and Katrina Ligett.
\newblock Routing without regret: On convergence to nash equilibria of
  regret-minimizing algorithms in routing games.
\newblock In {\em Proceedings of the twenty-fifth annual ACM symposium on
  Principles of distributed computing}, pages 45--52, 2006.

\bibitem[BK{\etalchar{+}}84]{benoit1984finitely}
Jean-Pierre Benoit, Vijay Krishna, et~al.
\newblock Finitely repeated games.
\newblock 1984.

\bibitem[Blu98]{blum1998line}
Avrim Blum.
\newblock On-line algorithms in machine learning.
\newblock {\em Online algorithms}, pages 306--325, 1998.

\bibitem[BM07]{blum2007learning}
Avrim Blum and Yishay Monsour.
\newblock Learning, regret minimization, and equilibria.
\newblock 2007.

\bibitem[CBL06]{cesa2006prediction}
Nicolo Cesa-Bianchi and G{\'a}bor Lugosi.
\newblock {\em Prediction, learning, and games}.
\newblock Cambridge university press, 2006.

\bibitem[CK11]{conitzer2011}
Vincent Conitzer and Dmytro Korzhyk.
\newblock Commitment to correlated strategies.
\newblock In {\em Proceedings of the Twenty-Fifth AAAI Conference on Artificial
  Intelligence}, AAAI'11, page 632–637. AAAI Press, 2011.

\bibitem[{\v{C}}LBA21]{Cerny2021}
Jakub {\v{C}}ern{\`y}, Viliam Lis{\`y}, Branislav Bo{\v{s}}ansk{\`y}, and
  Bo~An.
\newblock Computing quantal stackelberg equilibrium in extensive-form games.
\newblock In {\em Proceedings of the AAAI Conference on Artificial
  Intelligence}, volume~35, pages 5260--5268, 2021.

\bibitem[CS06]{conitzer2006computing}
Vincent Conitzer and Tuomas Sandholm.
\newblock Computing the optimal strategy to commit to.
\newblock In {\em Proceedings of the 7th ACM conference on Electronic
  commerce}, pages 82--90, 2006.

\bibitem[FS99]{freund1999adaptive}
Yoav Freund and Robert~E Schapire.
\newblock Adaptive game playing using multiplicative weights.
\newblock {\em Games and Economic Behavior}, 29(1-2):79--103, 1999.

\bibitem[Hic35]{Stack35}
J.~R. Hicks.
\newblock {\em The Economic Journal}, 45(178):334--336, 1935.

\bibitem[HLNW22]{haghtalab2022learning}
Nika Haghtalab, Thodoris Lykouris, Sloan Nietert, and Alexander Wei.
\newblock Learning in stackelberg games with non-myopic agents.
\newblock In {\em Proceedings of the 23rd ACM Conference on Economics and
  Computation}, pages 917--918, 2022.

\bibitem[HPS14]{halpern2014truth}
Joseph~Y Halpern, Rafael Pass, and Lior Seeman.
\newblock The truth behind the myth of the folk theorem.
\newblock In {\em Proceedings of the 5th conference on Innovations in
  theoretical computer science}, pages 543--554, 2014.

\bibitem[Kar72]{karp1972reducibility}
Richard~M Karp.
\newblock Reducibility among combinatorial problems.
\newblock In {\em Complexity of computer computations}, pages 85--103.
  Springer, 1972.

\bibitem[KDB14]{krichene2014convergence}
Walid Krichene, Benjamin Drighes, and Alexandre Bayen.
\newblock On the convergence of no-regret learning in selfish routing.
\newblock In {\em International Conference on Machine Learning}, pages
  163--171. PMLR, 2014.

\bibitem[KM20]{Karwowski2020}
Jan Karwowski and Jecek Ma{\'{n} }dziuk.
\newblock Double-oracle sampling method for stackelberg equilibrium
  approximation in general-sum extensive-form games.
\newblock {\em Proceedings of the {AAAI} Conference on Artificial
  Intelligence}, 34(02):2054--2061, apr 2020.

\bibitem[LC10]{Letchford2010}
Joshua Letchford and Vincent Conitzer.
\newblock Computing optimal strategies to commit to in extensive-form games.
\newblock In {\em Proceedings of the 11th ACM Conference on Electronic
  Commerce}, EC '10, page 83–92, New York, NY, USA, 2010. Association for
  Computing Machinery.

\bibitem[LGH{\etalchar{+}}22]{lauffer2022}
Niklas Lauffer, Mahsa Ghasemi, Abolfazl Hashemi, Yagiz Savas, and Ufuk Topcu.
\newblock No-regret learning in dynamic stackelberg games, 2022.

\bibitem[LMC{\etalchar{+}}12]{letchford2012}
Joshua Letchford, Liam MacDermed, Vincent Conitzer, Ronald Parr, and Charles~L.
  Isbell.
\newblock Computing stackelberg strategies in stochastic games.
\newblock {\em SIGecom Exch.}, 11(2):36–40, dec 2012.

\bibitem[LS05]{littman2005polynomial}
Michael~L Littman and Peter Stone.
\newblock A polynomial-time nash equilibrium algorithm for repeated games.
\newblock {\em Decision Support Systems}, 39(1):55--66, 2005.

\bibitem[MTS12]{Janusz2012}
Janusz Marecki, Gerry Tesauro, and Richard Segal.
\newblock Playing repeated stackelberg games with unknown opponents.
\newblock In {\em Proceedings of the 11th International Conference on
  Autonomous Agents and Multiagent Systems - Volume 2}, AAMAS '12, page
  821–828, Richland, SC, 2012. International Foundation for Autonomous Agents
  and Multiagent Systems.

\bibitem[Rub16]{rubinstein2016settling}
Aviad Rubinstein.
\newblock Settling the complexity of computing approximate two-player nash
  equilibria.
\newblock In {\em 2016 IEEE 57th Annual Symposium on Foundations of Computer
  Science (FOCS)}, pages 258--265. IEEE, 2016.

\bibitem[S{\etalchar{+}}03]{schrijver2003combinatorial}
Alexander Schrijver et~al.
\newblock {\em Combinatorial optimization: polyhedra and efficiency},
  volume~24.
\newblock Springer, 2003.

\bibitem[SC73]{Simaan73}
M.~Simaan and J.~B. Cruz.
\newblock On the stackelberg strategy in nonzero-sum games.
\newblock {\em J. Optim. Theory Appl.}, 11(5):533–555, may 1973.

\bibitem[vBA20]{cerny2020}
Jakub \v{C}ern\'{y}, Branislav Bosansk\'{y}, and Bo~An.
\newblock Finite state machines play extensive-form games.
\newblock In {\em Proceedings of the 21st ACM Conference on Economics and
  Computation}, EC '20, page 509–533, New York, NY, USA, 2020. Association
  for Computing Machinery.

\bibitem[VSZ10]{von2010Leadership}
Bernhard Von~Stengel and Shmuel Zamir.
\newblock Leadership games with convex strategy sets.
\newblock {\em Games and Economic Behavior}, 69(2):446--457, 2010.

\bibitem[ZT15]{Zuo2015OptimalMS}
Song Zuo and Pingzhong Tang.
\newblock Optimal machine strategies to commit to in two-person repeated games.
\newblock In {\em AAAI}, 2015.

\end{thebibliography}
\bibliographystyle{alpha}

\appendix

\section{Supplementary Related Work}
\label{app:rw}

In~\cite{conitzer2011}, Conitzer and Korzhyk observe that instead of solving a linear program for each of the $k$ pure strategies to find the SE, it is possible to instead solve a single linear program over the $k^{2}$ move pairs. This approach gives equivalent solutions for the two-player setting, but, when armed with a correlating device, allows them to compute optimal \emph{correlated} strategies to commit to in three-player games. The problem of optimizing over distributions of move pairs is technically similar to our own LP, though the two LPs have different constraints as they have differing goals.
Conitzer and Korzhyk's linear program is used to find an optimal correlated 
equilibria to commit to, which could involve randomness from both parties, while our LP has the goal 
of finding a hypothetical ideal distribution over move pairs that can be stably enforced by a particular kind of threat in a repeated two player game with a Leader.

Commitment and correlation are further explored in~\cite{letchford2012}, which gives some hardness results for computing SE in two-player infinite-horizon stochastic games. They prove NP-hardness for exact Stackelberg in the case where the Leader is allowed to commit to a mixed strategy at each stage. Any finite-horizon repeated game can be represented as an infinite-horizon stochastic game by having $T$ states which always transition in line and a $(T+1)$-th state with a self-loop which has payoff $0$ for both players. However, infinite-horizon stochastic games are a strictly richer class of games, and therefore this hardness result does not imply any complexity results for our case.~\cite{letchford2012} then show a poly-time algorithm to find an approximate Stackelberg for the case where the Leader is allowed to send signals to the Follower throughout the game, which allows them to commit to a correlated equilibria. In our setting, this would be equivalent to allowing the Leader, at round $t$, to send not only their move but also information about the realized randomness of a future move. 
Our Leader algorithms cannot send any information other than the moves at each round, and therefore this poly-time approximation does not imply any complexity result for our setting. 

Another line of work adjacent to finite-horizon repeated games is that of finding SE in extensive-form games (EFGs). EFGs can be represented as trees, with paths down the tree representing move sequences and leaves representing outcomes. Here, a Stackelberg strategy is usually presented as a mapping from each node in the game tree to a mixed strategy.~\cite{Letchford2010} show that computing SE for EFGs with stochastic transitions is NP-hard in the size of the game tree. There exist exact(\cite{cerny2020}) and approximate (\cite{Karwowski2020},~\cite{Cerny2021}) algorithms for computing SE for various special cases of the EFG setting, all of which require time linear or polynomial in the size of the game tree. Most relevantly to our work,~\cite{Letchford2010} find an algorithm to compute Stackelberg strategies for the special case of complete-information, two-player determinstic EFGs that is polynomial in the size of the game tree.~\cite{Bosansky2017} improve this to an algorithm which is linear in the game tree size. Finite-horizon repeated games can themselves be seen as a special case of complete-information, deterministic EFGs, and these techniques can therefore be used in our setting. However, the size of a game tree in a repeated game is exponential in $T$ --- note that on each of the $T$ rounds, there are $n^2$ different possible outcomes. Therefore, the pre-existing algorithms for EFGs imply only an exponential-time algorithm for repeated games. Our work goes beyond this to find an approximation algorithm for repeated games which is polynomial in $T$ and in the size of the single-shot game, and thus sublinear in the size of the corresponding game tree. 

\section{Proofs from Section~\ref{sec:prelims}}

\subsection{Proof of Lemma~\ref{lemma:bre}}
\label{app:bre}

Let us define the set $S$ to be the set of all deterministic $\Gpa$s which pick output strategies for each input via a lookup table~\footnote{The emphasis on lookup tables is to have a canonical form for deterministic $\Gpa$s so as to prevent redundancies.}. Using our precise definition of $\Gpa$s, a $\Gpa$ is in $S$ iff it is comprised of $T$ finite-time algorithms, where the $i$th algorithm takes as input a history of play of the past $i-1$ rounds and directly uses their lookup table to find what move to output. Note that, as none of the algorithms take random bits as input, no coin flips can be performed and Player 2 is always playing a pure strategy. This class of $\Gpa$s is therefore parameterized entirely by a lookup table. There are $n^{2(t-1)}$ possible transcripts of play that could be inputs to the table, and $n$ possible pure strategies that could be outputs. Thus there are $n^{n^{2(t-1)}} \leq n^{n^{2T}}$ different possible $f_{t}$. Every $\Gpa$ has $T$ such algorithms. Thus we can upper bound the number of $\Gpa$s in $S$ by $n^{n^{2T}T}$. Therefore, $S$ is a finite set. 

Now we will prove that, for any Leader $\Gpa$, the Follower has a best response that lies in $S$. We will proceed by backwards induction, proving the claim that, beginning at any time $i$, the Follower has a response maximizing their payoff for the remainder of the game which uses no randomness. Additionally, we will also show (as part of the same induction hypothesis) that among all best responses by the Follower, a best response that uses no randomness beginning at time $i$ is in fact optimal for the Leader.  

Base case: At round $T$, regardless of the previous transcript of play, there exists some deterministic strategy for Player 2 which maximizes payoff for the remainder of the game. Since this is the final round, Player 1 can no longer adapt to Player 2's moves. Therefore Player 2 can do no better than maximizing their payoff, and maximizing Player 1's payoff subject to the same. Player 2 has access to the Leader's Algorithm at round $T$, called $f_{T}$, to the transcript of play, and to how Player 1 utilizes randomness in $f_{1}...f_{T-1}$. Player 2 can therefore compute the distribution over coins which are used in $f_{T}$. If it is a coin that has never been used before, then the probability of heads is $\frac{1}{2}$. if it is a coin that has been used in earlier function(s), then the probability of heads is conditioned on the realizations of those previous moves. Using this computed distribution on coin flips, Player 2 can express Player 1's strategy at round $T$ as a mixed strategy. It is well known that there exists a pure strategy which is an optimal response in expectation to any mixed strategy. Therefore, Player 2 can play a pure strategy at round $T$ which maximizes payoff for the remainder of the game (and also maximizes Player 1's payoff subject to the above). 

Inductive hypothesis: If there exists some deterministic $\Gpa$ for Player 2 which maximizes their payoff starting from round $t$, there exists some deterministic $\Gpa$ which maximizes their payoff starting from round $t-1$. 
Proof: As in round $T$, Player 2 can compute a probability distribution $s_{1}$ over Player 1's possible moves at round $t-1$. Using the inductive hypothesis, for any possible strategy they can play at time $t-1$ and realization of Player 1's move, they can also compute a non-random optimal strategy for rounds $t$ to $T$. The expected payoff of Player 2 from round $t-1$ to $T$, given they play a (possible mixed) strategy $s_{2}$ at round $t-1$, will be their expected payoff at round $t-1$, plus the expected Follower payoff $F(O_{t \rightarrow T})$ of their optimal strategy $O_{t \rightarrow T}$ going forward, conditioned on $s_{2}$ and $s_{1}$. Let $s^{*}_{1}$ and $s^{*}_{2}$ be the realizations of the distributions. Note that they are conditionally independent for a fixed $s_{1}$ and $s_{2}$. Then, the expected payoff of Player 2 can be expressed as 
$$\sum_{j=1}^{n}\sum_{i=1}^{n}(M_{2}(i,j)+ \E[F(O_{t\rightarrow T})|s_{1}^{*} = i, s_{2}^{*} = j]) P(s_{1}^{*}=i)P(s_{2}^{*}=j)$$

The outer summation is simply a weighted sum of expected payoffs for all pure strategies in the support of $s_{2}$. Thus, there is some particular pure strategy $s^{*}_{2}$ for which $\sum_{i=1}^{k}(M_{2}(i,j)+ E[O_{t\rightarrow T}|s_{1}^{*} = i, s_{2}^{*} = j])P(s_{1}^{*}=i)$ is at least as large as the expected payoff using $s_{2}$. Therefore, it is optimal for Player 2 to play this pure strategy in the $t$-th round followed by the optimal deterministic $\Gpa$ starting at round $t$, extending the induction hypothesis. 

A similar argument can be used to extend the induction hypothesis to include that there exists an optimal (for Player 1) best response (for Player 2) that is deterministic --- replacing $M_2$ by $M_1$ allows one to see this. More explicitly, we have in fact shown that any best response for the Follower must necessarily play one among a set of actions $S$ in the $(t-1)$-th round. Therefore, we can think of any best response for the Follower as playing a distribution $P(S)$ over actions in $S$ in round $t-1$. By the induction hypothesis, we know that a Leader-optimal best response for the Follower starting at the $t$-th round can be deterministic. Therefore, once the action $j \in S$ played by the Follower in round $t-1$ is fixed, for each action $i$ played by the Leader, $O_{t \rightarrow T}$ is a deterministic Leader optimal best response for the Follower, with associated payoff $L(O_{t \rightarrow T})$. Thus, the payoff of the Leader is 

$$\sum_{j \in S}\sum_{i=1}^{n}(M_{1}(i,j)+ \E[L(O_{t\rightarrow T})|s_{1}^{*} = i, s_{2}^{*} = j])P(s_{1}^{*}=i)P(s_{2}^{*}=j)$$

Using a similar argument as the one above allows us to conclude that among the set of best responses, there exists one that is optimal for the Leader and is deterministic in round $t-1$, thus extending the induction hypothesis.



\subsection{Proof of Theorem~\ref{thm:compact_existence}}
\label{app:compact}

Let us assume that our algorithms can draw arbitrarily precise samples from any real valued distribution. We will show that this assumption is without loss of generality, even though we only access to a finite number of random coins.

We begin our proof by observing that every $\Gpa$ can be rewritten as a probability distribution over $\mathcal{L}$, the set of deterministic look-up table $\Gpa$s. Since $\mathcal{L}$ is a finite sized set (at most $n^{n^{2T} T}$), this is a finite support probability distribution. Consequently, the set of $\Gpa$s can be seen to be exactly the probability simplex $\Delta^{k-1} \in \mathbb{R}^k$ where $k = |\mathcal{L}|$~\footnote{The assumption made at the beginning is key to this statement}. To see this, observe that once all the randomness is realized, we are left with a deterministic $\Gpa$, allowing us to construct a measure over the set of deterministic $\Gpa$s. Consequently, the repeated game can now be viewed as a (large) normal form game where the set of actions for each player is the respective set of deterministic $\Gpa$s. Therefore, the result of Conitzer and Sandholm~\cite{conitzer2006computing} implies the existence of a Stackeleberg Equilibrium for the game. Even more specifically, their algorithm, built around picking the best solution out of a finite set of linear programs, implies that there is a optimal distribution for the Leader to play that uses rational weights with a bounded number of bits needed to represent them. Therefore, we can in fact find a Stackelberg $\Gpa$ for the Leader that uses only a finite number of random coins, justifying our original assumption to be without loss of any generality.

\subsection{Proofs from Section~\ref{sec:separation}}
\label{app:gap_proofs}
We reproduce the $\mathcal{PD}$ matrix here:
\[ \begin{bmatrix}
(3,3) & (0,5) \\
(5,0) & (1, 1)
\end{bmatrix} \]

The Prisoner's Dilemma game is a classic illustration of why rational players may fail to cooperate despite the fact that doing so would lead to mutually higher payoffs. The only Nash equilibrium is both players defecting and picking up payoffs $(1,1)$, even though both players would do better by both cooperating and getting $(3,3)$. These same dynamics remain at play in the single-round Stackleberg equilibrium because defecting is a strictly dominant strategy for both players. Thus, regardless of the strategy that the Leader declares, the Follower's best response is always to defect. The Leader can therefore do no better than to commit to defecting themselves and securing a payoff of $1$. 

In a $T$-round repeated game, the Leader could simply repeat this strategy to garner a payoff of $T$. However, they can perform much better by leveraging the multiple rounds of play. Consider the following Leader $\Gpa$: cooperate in every round until the first round that the Follower defects. If the Follower ever defects, the Leader will defect every round for the remainder of the game. We show that the Follower's best response $\Gpa$ is to cooperate for the first $T-1$ rounds, garnering the Leader a payoff of $3(T-1)$. 

The Follower's best response $\Gpa$ is to cooperate for the first $T-1$ rounds and defect only on the last round, netting the Follower a payoff of $3 \cdot (T-1) + 5 = 3T + 2$. To prove this is the best response $\Gpa$, consider any other deterministic $\Gpa$ for the Follower (through the proof of Lemma~\ref{lemma:bre}, we can restrict the Follower's selection to deterministic $\Gpa$s). If the Follower $\Gpa$ never has the possibility of defecting at any round, the Follower value is upper bounded by $3T$ and is hence not a best response. Now consider any $\Gpa$ for the Follower that results in column 2 being played for the first time at round $t \leq T-1$. 
The resulting payoff for the Follower in such an event is upper bounded by $3 \cdot (t-1) + 5 + 1 \cdot (T-t) = T + 2t + 2$. As $t \leq T-1$, this is at most $T + 2(T-1) + 2 = 3T$. This is again less than the response $\Gpa$ described above, completing the optimality proof. 

The induced transcript of the game will have both players cooperating until the last round, where the Leader cooperates and the Follower defects. This will get the Leader a payoff of $3(T-1)$, which for any $T>1$ is strictly better than the static Stackelberg strategy. 
 
For any $T>1$, this is strictly better than repeating the single-round Stackelberg strategy $T$ times. As there exists some multi-round $\Gpa$ with a gap in payoff, certainly the multi-round Stackelberg $\Gpa$ will have at least as large a gap.

We have now established that the finite-horizon repeated game Stackelberg $\Gpa$ is no simple extension of the static Stackleberg strategy. Furthermore, this example gives us some intuition for what good Leader $\Gpa$s might look like in general. The Leader was able to leverage a threat (defecting for the remainder of the game) in order to force the Follower to obey a specific move sequence (Follower cooperates, Leader cooperates). While implementing this threat would harm both the Leader and the Follower, the best-responding Follower does better by obeying the Leader than by being exposed to the threat, guaranteeing that they will follow this sequence. Consequently, the threat will in fact never be implemented, allowing the Leader to reap the rewards of the intended sequence. 

In the above example, the power of multi-round commitments allowed the Leader to force a cooperation strategy which benefited both players approximately equally. This is similar in spirit to results about Nash Equilibria in infinite-horizon repeated Prisoner's Dilemma games. However, the rich space of $\Gpa$s available to the Stackelberg Leader in different games includes many powerful Leader strategies which are not in their nature mutually beneficial. Often, the Leader can do better than they would in any coordination strategy at the  expense of the Follower. Consider a different Leader $\Gpa$ for the same Prisoner's Dilemma game, with two phases: the Leader will defect for the first $\frac{T}{2}$ rounds of play, and cooperate for the final $\frac{T}{2}$ rounds of play. However, if the Follower ever defects in either phase, the Leader will immediately switch to defecting for the remainder of the game. Using a similar analysis as before, we can see that the Follower's best response is to defect only on the last round, getting them a payoff of $3 \cdot (\frac{T}{2} - 1) + 5 = \frac{3T}{2} + 2$. The Leader will now get an improved payoff of $5 \cdot \frac{T}{2} + 3 \cdot (\frac{T}{2}-1) = 4T - 3$. 

The Leader was able to extract more value from the game in this new $\Gpa$ because, even though this $\Gpa$ is worse for the Follower than the previous $\Gpa$, playing along with the Leader still gets the Follower a higher payoff than disobeying. It is possible for the Leader to push it too far, however--if the Leader makes the first phase (in which they defect) last $T-1$ rounds, for example, then the Follower does better by immediately defecting than by staying silent. Then the Leader does no better than they do in the static Stackleberg strategy. One might wonder exactly how far the Leader can push the envelope and still expect the Follower to obey. In section~\ref{sec:main_alg}, we describe an LP which solves this sort of optimization problem, and prove that Stackleberg $\Gpa$s constructed this way are optimal. Using the ideas shown there, we can explicitly construct an approximately optimal Stackelberg strategy for the finite-horizon repeated Prisoner's Dilemma game. This construction is shown in Appendix~\ref{app:pd_construction}.

One may observe that while the Follower's payoff here is still higher than in the single-round Stackleberg game (where they get value $T$). This will always be true for Prisoner's Dilemma Games, as the Follower can always guarantee themselves their equilibrium value no matter what the Leader does. However, it is not generally the case that repeated games benefit all parties involved; in fact, there exist simple two player repeated games in which the Follower gets arbitrarily worse payoff on average than in the single-shot case when responding to an optimal Stackelberg Leader.


\subsection{Separation between Learning and Full Information}
\label{app:learning_gap}

\begin{lemma}
\label{lemma:learning_gap}
There exist static repeated games such that a Stackelberg Leader in the learning setting can get a payoff no greater than $\frac{S}{2 - \epsilon}$, where $S$ is the single-round Stackelberg value and $0 < \epsilon < 2$.
\end{lemma}

We know that the Leader can always guarantee themselves the stage game Stackelberg value in the full information setting. Therefore, an immediate corollary of the lemma above is that, in some games, the Stackelberg value in the full information setting is strictly higher than in the learning setting.



\begin{proof}

Let us assume for contradiction that there exists some algorithm $A$ which takes as input a time horizon $T$ and outputs a GPA that gets at least a $\frac{1}{2-\epsilon}$-approximation to the static Stackelberg value on average. The algorithm does not take the matrix as input, but the GPA can be adaptive to the opponent's moves and the payoffs that the Leader gets each round.

Now, consider the following matrix $M1$, where the Leader plays the rows and the Follower plays the columns:
\[ \begin{bmatrix}
(1,\frac{\epsilon}{2}-1) & (0,0) \\
(0,\frac{\epsilon}{2}) & (0, 0)
\end{bmatrix} \]

Here, the static Stackeleberg value is $\frac{\epsilon}{2}$. So the algorithm must get the Leader a value of at least $\frac{\epsilon}{2(2 - \epsilon)}$. Consider some optimal GPA Follower response to $M1$, and call it $G_{1}$. The induced sequence of play might feature randomness in each round from both the Leader and the Follower. We aggregate the probability of each action pair over all the rounds of play, divide by the number of rounds $T$, and refer to this quantity as the frequency of the action pair.
We observe that $L1F1$ (the first action for the Follower and the first action for the Leader) must have frequency at least $\frac{\epsilon}{2(2 - \epsilon)}$. The Follower could always get $0$ by playing their second action each round, and then $L1F1$ would never be played. Therefore, in order for $G_{1}$ to be a best response, the Follower's average payoff overall must be at least $0$. But every time that $L1F1$ is played they get a payoff of $\frac{\epsilon}{2} - 1$. They only get a positive payoff with move pair $L1F2$, when they get $\frac{\epsilon}{2}$. Thus:
$$P(L1F1)\cdot \left(\frac{\epsilon}{2} - 1\right) + P(L1F2)(\frac{\epsilon}{2}) \geq 0$$
$$P(L1F1)\cdot (\epsilon -2) + P(L1F2) \cdot \epsilon \geq 0$$
$$P(L1F2)  \geq \left(\frac{2 - \epsilon}{\epsilon}\right) P(L1F1) \geq \left(\frac{2 - \epsilon}{\epsilon}\right)  \cdot \frac{\epsilon}{2(2 - \epsilon)} = \frac{1}{2}$$

$$P(L1F2) \geq \frac{1}{2}$$

Now, consider the following matrix $M2$, which has the same payoffs for the Leader as $M1$:
\[ \begin{bmatrix}
(1,0) & (0,0) \\
(0,1) & (0, 0)
\end{bmatrix} \]

Here, the static Stackelberg value is $1$. So the algorithm must get the Leader a value of at least $\frac{1}{2 - \epsilon}$. This means that they must play move $L1$ while the Follower plays move $F1$ with frequency at least $\frac{1}{2 - \epsilon}$ on average. I.e., $L1F1$ has that frequency. Now, consider some optimal GPA Follower response to $M2$, and call it $G_{2}$. Since $L1F1$ has frequency at least $\frac{1}{2 - \epsilon}$, and the Follower gets payoff $0$ in each of these rounds, their total expected payoff is upper bounded by $1 - \frac{1}{2 - \epsilon} = \frac{1 - \epsilon}{2 - \epsilon}$.  

Now, consider an alternate GPA for this Follower: play $G_{1}$. Note that $M1$ and $M2$ have the same payoffs for the Leader at every move, and the same number of moves. So if a Follower plays the same algorithm on $M1$ and $M2$, the Leader's algorithm cannot distinguish between the matrices and therefore must behave identically. So, when playing $G_{1}$, $L1F2$ must still have frequency at least $\frac{1}{2}$. Whenever this move pair is played, the Follower gets payoff $1$. This means that the Follower gets total expected payoff at least $\frac{1}{2}$. This is greater than $\frac{1 - \epsilon}{2 - \epsilon}$, so $G_{2}$ is not a best response to $M2$. Thus we have derived a contradiction.  

\end{proof}

\section{General Hardness Of Finding Optimal Response $\Gpa$s}
\label{app:br_hard}

When reasoning about the computational complexity of $\Gpa$s, we must distinguish between two types of computational costs: the complexity of constructing a $\Gpa$, and the complexity of actually running that $\Gpa$ to generate moves. In the previous sections we have not placed any constraints upon the computational complexity of the Leader and Follower $\Gpa$s, other than to specify that all of their $\Gpa$ functions must run in finite time. Furthermore, we have given the Follower arbitrary computational power to construct their best response $\Gpa$ after seeing the Leader $\Gpa$. Despite the rich space of potential $\Gpa$s that this definition allows, we have shown the existence of an approximate Stackelberg $\Gpa$ for any game which is computationally efficient in all possible notions of the word; not only does the Leader's $\Gpa$ run in polynomial time, the Follower can find a poly-time best response to it in polynomial time (Lemma~\ref{lemma:br_poly}). 
Thus, our LP-inspired $\Gpa$ is not only approximately optimal in the general setting, but also offers a realistic prediction since both the Leader as well as the Follower can convince themselves of the (near) optimality of their respective $\Gpa$s. 

This simplicity should not be taken for granted. In general, the problem of finding an efficient best response $\Gpa$ is NP-hard. This hardness does not simply stem from the potential for the Leader to construct inscrutable $\Gpa$s; in fact, there exist poly-time, succinctly representable Leader $\Gpa$s for which the Follower cannot find a poly-time best response, unless P=NP. We give an example of such a $\Gpa$ in Lemma~\ref{lemma:br_hard}. To make this problem well defined, we allow the Follower's $\Gpa$ to take as input some representation of the Leader's $\Gpa$ (in this case a polynomial-time, polynomial-size Turing Machine) and analyze its run-time in terms of the size of the Leader's $\Gpa$ and the size of the game. In particular, we restrict the Leader's $\Gpa$ to a simple Turing Machine that checks if a given coloring of a graph is valid (showing that the complexity does not stem from any sophistication in the Leader's $\Gpa$).

\begin{lemma}
\label{lemma:br_hard}
Computing a set of polynomial-time Turing Machines (along with a suitable set of random coins) that constitute an optimal Follower response is NP-hard, even when the Leader's $\Gpa$s are constrained to polynomial-time computable functions with poly-size representations. 
\end{lemma}  
{\bf Note: } Here, when we say polynomial time, we mean polynomial time in the size of the game and $T$.\\

The proof below also implies the following lemma as well:
\begin{lemma}
\label{lemma:value_hard}
Given a Leader $\Gpa$ which is composed of poly-time computable functions with a poly-size representation, determining the expected value of the $\Gpa$ against a best-responding Follower is NP-hard.
\end{lemma}  
\begin{proof}
Assume that we have some algorithm $A(P, M, T)$ for the Follower which, given any poly-time Leader $\Gpa$ $\pol_{l}$, matrix $M$ and time horizon $T$, computes some $P_{f}$ which runs in polynomial time. We will show that $A$ is NP-hard via a reduction from the minimization version of graph coloring.

Consider any graph $G$ with $n$ nodes. The problem is to find the smallest number of colors needed to color $G$. To do this, we will construct the following game matrix $M$ and $\Gpa$ $\pol_{l}$ as input to $A$:

$M$ is an $n$ by $n$ game matrix for which $\mu_{2}(i,j) = 0$ for all $i,j \neq n$, and $\mu_{2}(n,n) = 1$. Let $T = n$.

Consider the following declared $\Gpa$ by Player 1, $P_{l}(G)$: 
$$f_{i} = 1 \forall i < T$$
$$\mathbb{P}(f_{T} = n) = \frac{1 - g(G,\transcript[0:T-1])}{n}$$
$$\mathbb{P}(f_{T} = 1) = \frac{g(G,\transcript[0:T-1])}{n}$$

$g(G,\transcript[0:T-1])$ is defined as follows:\\

Consider all of Player 2's moves at time $1:T-1$. If they ever play their $n$th move, terminate and return $n$. Otherwise, consider the move played at time $t$ to be a coloring for the $t$th node in $G$. As there are $T-1$ rounds to represent all vertices, and $n-1 = T-1$ valid moves by Player 2 to represent the maximum number of colors needed, Player 2's move sequence in the first $T-1$ rounds may represent any coloring of the graph. Check whether this represents a valid coloring in $O(n^{2})$ time by checking that Player 2 never played the same move on rounds representing adjacent vertices. If the coloring is invalid, return $n$. If the coloring is valid, return the number of unique moves used by Player 2. This will occur in polynomial time, and, in addition, the $\Gpa$ can be represented in a polynomial number of bits.

To find the minimum coloring, we will compute $P_{f} = A(P_{l}(G), M, n)$. Then, we will simulate running $P_{l}$ against $P_{f}$ to get a transcript. This will occur in polynomial time, as both $P_{f}$ and $P_{l}$ are poly-time. Then, we will return the total number of unique moves by Player 2 in the first $T-1$ rounds as the minimum coloring. \\

Correctness:
Consider the optimal strategy for Player 2 given this $\Gpa$. Their payoff is always $0$ unless Player 1 plays their $n$th move in the final round. At the final round, Player 2 can never do better than playing their $n$th move in the hope of getting some nonzero payoff. Thus, Player 2's expected payoff for using Follower $\Gpa$ $\pol_{f}$ is 
$$1 \cdot \mathbb{P}(f_{T} = n | \pol_{f}) = \frac{1 - \E[g_{G}(\transcript[0:T-1])|\pol_{f}]}{n}$$

Therefore the $\Gpa$ which maximizes Player 2's payoff is the $\Gpa$ $\pol_{f}$ which minimizes $$\E[g_{G}(\transcript[0:T-1])|\pol_{f}]$$

Note that as $c$ is the smallest number of colors that are needed to color $G$, there is no way for Player 2 to achieve a smaller value than $c$. This is because the move sequence for Player 2 in the first $t-1$ rounds must represent a valid coloring, and there is no valid coloring using less than $c$ colors. In addition, all Follower $\Gpa$s which achieve a value of $c$ must represent min colorings of $G$, followed by their $n$th move. Therefore, the first $T-1$ rounds are a minimum coloring. Furthermore, the number of unique moves used in these rounds is the number of unique colors used in the min coloring. Therefore, this reduction correctly outputs the min coloring value. As we have reduced from an NP-hard problem to $A$, $A$ is NP-hard. 
\end{proof}

\begin{lemma}
\label{lemma:br_poly}
If the  Leader's declared $\Gpa$ is a candidate (approximate) Stackelberg $\Gpa$ constructed by either of the approaches in Theorem~\ref{thm:stack_policy}, then there exists a linear-time algorithm for Player 2 to find an optimal response $\Gpa$. 
\end{lemma} 

\begin{proof}
The Follower can use the following algorithm to find their best-response $\Gpa$: check if deterministically playing the moves prescribed to them by the Leader's $\Gpa$ is optimal. If yes, output the $\Gpa$ doing exactly this, and if no, output some arbitrary deterministic $\Gpa$. By Lemma~\ref{lemma:follow_presc}, as long as the Leader's $\Gpa$ is constructed as per either of the algorithms in Theorem~\ref{thm:stack_policy}, playing the moves prescribed is optimal, so the verification will always result in the yes instance. Now we only need to prove that it is possible to perform this verification in polynomial time in $n$ and $T$. Similar to the spirit of the proof of Lemma~\ref{lemma:follow_presc}, the Follower can proceed by induction on the number of rounds, proving to themselves that there is no point at which they would prefer to deviate. This will take linear time in $n$ and $T$.
\end{proof}

\section{Proof from Section~\ref{sec:main_alg}}
\label{app:main_alg}

\subsection{Proof of Lemma~\ref{lemma:lp_upper_bd}}
\label{app:lp_upper_bd}


Assume for contradiction that for $M$, there exists a $\Gpa$ $L^{*}$ that the Leader can declare such that, when the Follower best responds, the expected average Leader payoff is $p^* > \lpopt$. Just like the Leader's $\Gpa$, the Follower's $\Gpa$ can be described as $T$ different functions, where the $t$-th function takes as input the transcript of play of length $t-1$ and all coin flips used by the Leader and outputs an action drawn from a probability distribution over pure actions to play in round $t$. So the Follower will best responding to $L^{*}$ with some $\Gpa$ $F^{*}$. Reasoning about the precise behavior of these $\Gpa$s in tandem might be complex--both $\Gpa$s could be randomized and could be sensitive to different realizations of their or their opponent's randomness. But, when played against each other, $(L^{*}, F^{*})$ will always produce an output which is a sequence of $T$ move pairs. Furthermore, this sequence is determined only by the declared $\Gpa$s and the coin flips of each player. Therefore, $(L^{*}, F^{*})$ induces a probability distribution over sequence outcomes. \\

We can think of this distribution as a vector of all possible $T$-length move pair sequences. $\mathbb{P}(S=s_i)$ is the weight in the distribution placed upon the $i$th sequence $s_i$ (for some arbitrary indexing of sequences). For the $i$-th sequence of move pairs, let $v_{1}(s_i)$ be the total payoff of the Leader and let $v_{2}(s_i)$ be the total payoff of the Follower. So, the expected payoff per round for the Leader is
$$p^* = \frac{1}{T} \sum_{i \in S}v_{1}(s_{i})\cdot \mathbb{P}(S = s_i)$$

$s_{i}$ is some particular sequence of move pairs $x_{i1},...,x_{iT}$, hence we can rewrite the value $p^*$ as

$$p^* = \frac{1}{T}\sum_{s \in S}\sum_{t=1}^{T} M_1(x_{st})\cdot \mathbb{P}(S=s_i)$$

Even within each sequence, the $x_{st}$'s are not necessarily unique, as a move pair can appear with some probability at different rounds and in different sequences. We can rewrite the sum by iterating over each unique move pair, which we index by $\{(i,j)\}$:

$$p^* = \frac{1}{T}\sum_{t=1}^{T} \left(\sum_{i,j}(M_1(i,j)\cdot \sum_{s \in S}\mathbb{P}(S = s_i, s[t] = (i,j)))\right)$$

For a particular round $t$, the sum of probabilities over all possible move pairs as part of each possible sequence adds up to 1, i.e., $\sum_{\forall i,j} \sum_{\forall s \in S}P(S = s_i, S[t] = (i,j)) = 1$. Therefore, for every fixed time $t$, there is a probability distribution over all move pairs. $p^*$ is thus simply the sum of $T$ different distributions over move pairs, divided by $T$. So $p^*$ can also be seen as $M_1$ applied to a distribution over move pairs. \\


Thus, we know that there exists some distribution over move pairs $\beta$ such that
$$p^* = \sum_{i,j}M_1(i,j)\beta_{i,j}$$

Using the same logic as above, just swapping $M_{2}$ for $M_{1}$, the value that the Follower gets in this game, which we will call $p_{F|L}$, can be written as 
$$p_{F|L} =  \sum_{i,j}M_2(i,j)\beta_{i,j}$$
As the Follower is best-responding, it must be the case that $p_{F|L} \geq V$. This is since by definition, the Follower can always guarantee themselves $V$ each round (for instance, by simulating the Leader's probability distribution in each round and picking a best-response pure strategy in expectation). 
Consequently, we have some distribution over move pairs $\beta$ such that 
$$\cdot \sum_{i,j}M_1(i,j)\beta_{i,j} = p^* > \lpopt$$
$$\cdot \sum_{i,j}M_{2}(i,j
)\beta_{i,j} \geq V$$
$$\sum_{i,j}\beta_{i,j}=1$$
$$\beta_{i,j} \geq 0$$

Thus, $\beta$ is a valid solution to the LP described in Section~\ref{sec:main_alg}, and the objective is larger than $\lpopt$. 
Therefore we have derived a contradiction, and there cannot exist a Leader $\Gpa$ $L^{*}$ that garners expected payoff above $\lpopt$.

\subsection{Proof of Lemma~\ref{lemma:follow_presc}}
\label{app:follow_presc}

We prove this by induction on the number of rounds that the Follower's $\Gpa$ follows the prescribed sequence. The base case is $t =0 $ and is trivially true. Assume that the Follower cooperates for the first $t$ rounds. Consider any deviation from the prescribed $\Gpa$ in the $t+1$-th round (where $t+1 <= c N$). The average payoff for the rounds $t+1$ to $c N$ on following the $\Gpa$ is at least the threat level $V$ --- this is because  $\sum_{k} M_2(k) \alpha_k \ge V$ and we are following a truncated version of the best part of this distribution (from the Follower's perspective) when the Follower cooperates with the prescribed $\Gpa$. The average payoff in the final $r >0$ rounds equals the maximum possible payoff $m$. Thus, the total expected payoff is at least $$V \cdot (c N - t) + r \cdot m = V \cdot (c N - t) + m \cdot (r-1) + m$$ Now, consider deviating in the $(t+1)$-th round. On this single round, the Follower may get value as high as $m$. In all remaining $T - t -1 = c \cdot N -t + r - 1$ rounds, the threat will be employed, and thus the Follower will get expected payoff no higher than $V$. So the expected payoff is at most
$$m + V \cdot (c \cdot N - t + r - 1) = V \cdot (c N - t) + V \cdot (r-1) + m$$

As $m \geq V$, the payoff of obeying the prescribed $\Gpa$ is at least as high as the payoff of deviating from the prescribed $\Gpa$. So the inductive step holds when $t \leq c \cdot N$. 

Finally, consider deviating on any round $t + 1 > c N$. By construction, for all remaining rounds the Leader prescribes only the action pair which gives the Follower their globally optimal payoff $m$. So, the remaining payoff by obeying is $m \cdot (T - t)$. As $m$ is the maximum payoff for the Follower, no other move sequence over the next $T-t$ rounds can get them a higher payoff. Thus, the best response by the Follower for all remaining rounds is to obey the prescribed sequence. So the inductive step also holds when $t > c \cdot N$. Therefore, the inductive step always holds, so a best-response $\Gpa$ for the Follower will always follow the prescribed sequence.

\subsection{Proof of Claim~\ref{claim:few_enough_swaps}}
\label{app:few_enough_swaps}

In the first case where $c \geq \frac{\sqrt{10}}{T'^{0.25}\sqrt{A}}$, let us upper bound the fraction of swaps needed (which we will write as $s$) by writing an expression for the final value for the Follower. As we swap starting with the lowest values for the Follower, their average payoff for the non-swapped action pairs will be at least their average payoff over all pairs before and swapping occurs. We know the final value just needs to be greater than $V$, so it is sufficient to have:

$$s\cdot m + (1 - s)\left(V - \frac{10}{\sqrt{T'}}\right) \geq V$$
Rewriting this:

$$sm  - sV + s\frac{10}{\sqrt{T'}} - \frac{10}{\sqrt{T'}} \geq 0$$

$$s(m - V) + s\frac{10}{\sqrt{T'}} - \frac{10}{\sqrt{T'}} \geq 0$$

$$s\left(c + \frac{10}{\sqrt{T'}}\right) - \frac{10}{\sqrt{T'}} \geq 0$$

$$s \geq \frac{10}{c\sqrt{T'} + 10}$$

Recall that $c \geq \frac{\sqrt{10}}{T'^{0.25}\sqrt{A}}$. So, the largest that $s$ needs to be is

$$s \geq \frac{10}{\frac{\sqrt{10}}{T'^{0.25}\sqrt{A}}\sqrt{T'} + 10}$$

Which is itself upper bounded by $$\frac{10}{\frac{\sqrt{10}}{T'^{0.25}\sqrt{A}}\sqrt{T'}} = \frac{\sqrt{10A}}{T'^{0.25}}$$

This is sublinear in $T'$. Thus, the average per-round loss to the Leader of this adjustment is at most $\frac{2\sqrt{10A}}{T'^{0.25}}$. 

Now, consider the other case: $c < \frac{\sqrt{10}}{T'^{0.25}\sqrt{A}}$

 We will now write an expression for the Follower value $V$ in the optimal LP. We will let $P_{m}$ represent the fraction of pairs that give the Follower value $m$. All other pairs give the Follower value at most $m - \frac{1}{A}$, by our definition of $A$. Thus, we have:

$$V = m-c = \leq P_{m}\cdot m + (1 - P_{m})\cdot \left(m - \frac{1}{A}\right)$$
Which we can solve to get

$$P_{m} \geq 1 - Ac$$

Recall that in this case, $c < \frac{\sqrt{10}}{T'^{0.25}\sqrt{A}}$. Thus, it must be that
$$P_{m} \geq 1 - \frac{\sqrt{10}A}{T'^{0.25}\sqrt{A}} = 1 - \frac{\sqrt{10A}}{T'^{0.25}}$$

Therefore, there are at most $\frac{\sqrt{10A}}{T'^{0.25}}$ fraction of pairs that do not give the Follower value $m$. Note that this is an upper bound on the fraction of pairs that must be swapped to achieve a value above $V$ for the Follower. Thus, we can upper bound $s$ to be $$s \leq \frac{\sqrt{10A}}{T'^{0.25}}$$

So the average additive loss (as compared to the empirical  distribution $\al$) contributed in either case case is at most $\frac{2\sqrt{10A}}{T'^{0.25}}$.

 We have seen that, with high probability, $\al$ achieves at most $\frac{10}{\sqrt{T'}}$ less value for the Leader than $\lpopt$. The analysis above also tells us that the Leader gets at most $\frac{2\sqrt{10A}}{T'^{0.25}}$ more under $\al$ as compared to $\al'$. 
Putting together the bounds from the analysis above, $\al'$ assures the Leader an average payoff that is off from $\lpopt$ by at most $\frac{3\sqrt{10A}}{T'^{0.25}}$.

\label{app:main_alg}

\subsection{Inevitability of Approximation to LP Upper Bound}
\label{app:inev}

One may wonder whether, using a more clever construction, it is possible to exactly achieve the $\lpopt$ value in every game and for every $T$. In fact, for many games the exact $\lpopt$ value is unachievable except in the limit. Below, we show an example game where the difference between the value of the LP and the average value of the best possible Leader GPA is $\frac{1}{T}$. For simplicity, we present $M_{1}$ and $M_{2}$ in a single matrix, with the Leader's payoff written first. The Leader is playing the rows and the Follower is playing the columns. 

\[ \begin{bmatrix}
(1,\frac{1}{2}) & (0,1) \\
(0,0) & (0,0)
\end{bmatrix} \]

Note that in this case the threat value used in the LP will be $0$, as if the Leader plays the second row the Follower will always get value $0$. Furthermore, the Leader only gets a payoff when move pair $(1,1)$ is played. So, the LP will simply solve the following maximization problem:

\begin{flalign*}
    \max 1 \cdot \alpha_{1,1} \qquad \qquad  \alpha \in \mathbb{R}^n &\\
    \text{ subject to } &\\
    \sum_{i,j} M_2(i,j) \alpha_{i,j} \ge 0 &\\
    \sum_{i,j} \alpha_{i,j} = 1 &\\
    \alpha_{i,j} \ge 0 \qquad \qquad \text{ for } i,j \in [n]^2 &
\end{flalign*}

Note that all $M_{2}(i,j) \geq 0$, and thus this LP is solved by setting $a_{1,1} = 1$ and getting value $1$.\\

Now, let us upper bound the best possible payoff that a Stackelberg GPA could achieve over $T$ rounds. Let's do this by examining the $T$-th round of play. Because it is the final round, the best that the Follower can do is maximize their expected payoff in this round. They will have the Leader's GPA in hand and thus will be able to calculate a probability distribution over the Leader's moves. If the Leader will play the first row with any nonzero probability, then a best responding Follower will always play the second column, giving the Leader a payoff of $0$. If the Leader will play the second row with probability $1$, then regardless of what the Follower does, the Leader will get a payoff of $0$. Thus, the Stackelberg GPA will always give the Leader a payoff of $0$ in the final round. The best that the Leader can do in the preceeding $T-1$ rounds is get their highest possible payoff of $1$. Therefore, their average payoff over all the rounds is upper bounded by

$$\frac{(T-1)\cdot 1 + 0}{T} = 1 - \frac{1}{T}$$

We have upper bounded the Stackelberg GPA value for this game to be $1 - \frac{1}{T}$, and shown that $\lpopt$ for this game is $1$. Therefore, we have lower bounded the gap between the Stackelberg GPA and $\lpopt$ to be $\frac{1}{T}$. 

\section{Proofs from Section~\ref{sec:hardness_three} }

\label{app:hardness_three}

\subsection{Proof of Lemma~\ref{lemma:first_half_hardness}}
\label{app:first_half_hardness}
If there exists a vertex cover of size at most $\frac{n}{2}$ in $G$, then consider the following strategies for Players 1 and 2 in a single round: Player 1 will pick a move uniformly at random which is associated with a vertex in the cover. Player 2 will pick a move uniformly at random which is associated with a vertex that is not in the cover. Now, consider the utility for Player 3 of playing any move $t_{v}$. Their payoff will be $0$ if the vertex they play is also played by either Player 1 or Player 2. By construction, every vertex in the graph is either being played by Player 1 w.p. $\frac{2}{n}$, or by Player 2 w.p. $\frac{2}{n}$. Therefore, their expected payoff for playing any $t_{v}$ will be $$\left (1 -\frac{2}{n} \right) \frac{n}{n-2} = 1$$

Furthermore, let us consider the utility for Player 3 of playing any move $t_{e}$. As long as Player 1 has played a move associated with an adjacent vertex to $e$, Player 3 will get value $0$. Because Player 1 is playing moves uniformly at random from a vertex cover of size $\frac{n}{2}$, there will be at least one vertex which is adjacent to every edge. So for any particular $t_{e}$, w.p. at least $1 - \frac{2}{n}$, the payoff will be $0$. Thus, Player 3's expected payoff for playing any $t_{e}$ will also be at most $\left (1 -\frac{2}{n} \right) \frac{n}{n-2} = 1$. Therefore, Player 3 has no moves that get them a payoff greater than $1$. Thus, playing $t_{0}$ is a best response, which gives all players an expected payoff of $1$. This is the optimal possible payoff for Player 2, and thus they are certainly best-responding to Player 1 here. Furthermore, this is the optimal possible payoff for Player 1, and thus we have a single-round SE where Player 1 gets value 1.

Now, observe the following GPA for Player 1: commit to playing this same distribution for all $T$ rounds. Player 2 best responds by committing to playing their corresponding half of the distribution for every round, because this will ensure that Player 2 gets their optimal payoff of $1$. Then, in each round, Player 3 is incentivized to play $t_{0}$, and everyone will get payoff $1$. This is in fact the optimal Stackelberg $\Gpa$ for Player 1 in this case, as they are getting on average the maximum possible value they get in the game. 

\subsection{Proof of Lemma~\ref{lemma:player3_min}}
\label{sec:player3_min}

We will proceed via proof by contradiction. Assume, for sake of contradiction, that there is no balanced vertex cover, but there exists some pair of mixed strategies by Players 1 and 2 such that Player 3 gets expected value no higher than $1 + \frac{1}{(n-2)n^{c-1}}$. Thus, in particular, there is no pure strategy $s$ that Player 3 can play which gets them greater than $1 + \frac{1}{(n-2)n^{c-1}}$. 

Consider taking any distribution by Player 1, and sorting the moves in their support in decreasing order of likelihood. Then, add all the moves not in the support at the end of the ordering. Let $p_{1}(x_{i})$ be the probability that Player 1 will play their $i$-th most likely move. We will now pay particular attention to their $(\frac{n}{2} + 1$)-th most likely move (WLOG, we assume that $n$ is even). Let us consider two cases:\\
1) $p_{1}(x_{\frac{n}{2}+1}) < \frac{1}{n} - \frac{1}{2n^{c}}$\\
2) $p_{1}(x_{\frac{n}{2}+1}) \geq \frac{1}{n} - \frac{1}{2n^{c}}$\\

Let us begin with case 1), i.e., $p_{1}(x_{\frac{n}{2}+1}) < \frac{1}{n} - \frac{1}{2n^{c}}$. Note that, because there is no vertex cover of size $\frac{n}{2}$, there is some edge $e$ which is not covered by any of the first $\frac{n}{2}$ vertices in the ordering. Additionally, by the structure of graphs, there are at most $2$ vertices which could cover it. By our case assumption, all remaining vertex-associated moves by Player 1 have probability of strictly less than $\frac{1}{n} - \frac{1}{2n^{c}}$. Thus, the highest probability with which Player 1 could play a vertex which covers edge $e$ is strictly less than $2 \cdot (\frac{1}{n} - \frac{1}{2n^{c}}) = \frac{2}{n} - \frac{1}{n^{c}}$. Therefore, if Player 3 plays $t_{e}$, their expected payoff is

$$\frac{n}{n-2} \cdot P(r_{v}\textit{ not adjacent to } t_{e}) + 0 \cdot P(r_{v}\textit{ adjacent to } t_{e})$$

$$> \frac{n}{n-2} \cdot \left(1 - \frac{2}{n} + \frac{1}{n^{c}} \right )$$

$$=  \frac{n-2}{n-2} + \frac{n}{n-2} \cdot  \frac{1}{n^{c}}$$
$$=  1 + \frac{1}{(n-2)n^{c-1}}$$

Thus, in case 1), there is a pure strategy $s$ which gets Player $3$ value greater than $=  1 + \frac{1}{(n-2)n^{c-1}}$. 

We now proceed to case 2), $p_{1}(x_{\frac{n}{2}+1}) \geq \frac{1}{n} - \frac{1}{2n^{c}}$. To prove a contradiction in this case, we use the following claim, the proof of which appears in Appendix~\ref{sec:inner_p3min}. The proof is based upon a careful water-filling argument that shows how to move probability mass around within Player 1' strategy while maintaining the assumed conditions as invariant.

\begin{claim}
\label{claim:inner_p3min}
If there is any pair of distributions that Players 1 and 2 can play where $p_{1}(x_{\frac{n}{2} + 1}) \geq \frac{1}{n} - \frac{1}{2n^{c}}$ 
such that Player 3 cannot get expected value strictly greater than $1 + \frac{1}{(n-2)n^{c-1}}$ 
then there is a (possibly different) pair of distributions $p'_1$ and $p'_2$ that Players 1 and 2 can play and some vertex $x_{i^*}$ where $p'_{1}(x_{i^*}) = \frac{1}{n} - \frac{1}{2n^{c}}$ 
such that Player 3 cannot get expected value strictly greater than $1 + \frac{1}{(n-2)n^{c-1}}$. 
\end{claim}

Thus, to prove that there must be a move that gets Player 3 expected value strictly greater than $1 + \frac{1}{(n-2)n^{c-1}}$, it is sufficient to prove that there must be such a move when $p_{1}(x_{i^*}) = \frac{1}{n} - \frac{1}{2n^{c}}$ for some vertex $x_{i^*}$. Assume, for sake of contradiction, that there is no such move. Then, it must be that the probability that either Player 1 or Player 2 (or both) play a move associated with any given vertex is at least 
$\frac{2}{n} - \frac{1}{n^{c}}$, for all vertices, since otherwise Player 3 could play a vertex move to get an expected payoff strictly larger than $1 + \frac{1}{(n-2)n^{c-1}}$. Let us first consider all vertices other than $x_{i^*}$, and call this set $V'$. Because all the moves are being played with probability $\frac{2}{n} - \frac{1}{n^{c}}$, the sum of the total probabilities (using the union bound vertex by vertex) that Players 1 and 2 place on $V'$ must be at least
$$(n-1) \left (\frac{2}{n} - \frac{1}{n^{c}} \right) = \frac{2(n-1)}{n} - \frac{n-1}{n^{c}}$$
By our assumption, Player 1 is playing $x_{i^*+1}$ with probability $\frac{1}{n} - \frac{1}{2n^{c}}$. Thus, the total probability that Player 1 places in $V'$ is $1 - \frac{1}{n} + \frac{1}{2n^{c}}$. Thus, Player 2's probability of playing a move in $V'$ must be at least

$$\frac{2(n-1)}{n} - \frac{n-1}{n^{c}} - \left (1 - \frac{1}{n} + \frac{1}{2n^{c}} \right )$$

Which means that Player 2's probability of playing $x_{i^*}$ must be at most 

$$1 - \left (\frac{2(n-1)}{n} - \frac{n-1}{n^{c}} - \left (1 - \frac{1}{n} + \frac{1}{2n^{c}} \right ) \right)$$
$$= 1 - \frac{2(n-1)}{n} + \frac{n-1}{n^{c}} + 1 - \frac{1}{n} + \frac{1}{2n^{c}}$$
$$= \frac{1}{n} + \frac{1}{n^{c-1}} - \frac{1}{2n^{c}}$$

Going back to our assumption, Player 1 has a probability of $\frac{1}{n} - \frac{1}{2n^{c}}$ of playing $x_{i^*}$. Recall that there is no coordination possible in the realizations of this randomness, so once the probabilities are set, their behavior is independent. So, we can calculate the total probability that this move is played by either or both players:

$$p_{1}(x_{i^*}) + p_{2}(x_{i^*}) - p_{1}(x_{i^*})p_{2}(x_{i^*}) = p_{1}(x_{i^*}) + p_{2}(x_{i^*}) (1-p_{1}(x_{i^*}) )$$

We show that this value is always less than $\frac{2}{n} - \frac{1}{n^{c}}$, and therefore Player 3 can play $t_{x_{i^*}}$ to get a sufficiently large payoff, resulting in a contradiction. 
We wish to prove the following inequality:

$$p_{1}(x_{i^*}) + p_{2}(x_{i^*}) - p_{1}(x_{i^*})p_{2}(x_{i^*}) < \frac{2}{n} - \frac{1}{n^{c}}$$

Since $p_{1}(x_{i^*})$ is constant and smaller than $1$, the left hand side is upper bounded by replacing $p_{2}(x_{i^*})$ by the maximum possible value it can take. So, we wish to prove the stronger inequality below, which we prove by rearranging terms:

$$\frac{1}{n} - \frac{1}{2n^{c}} + \frac{1}{n} + \frac{1}{n^{c-1}} - \frac{1}{2n^{c}} - \left(\frac{1}{n} - \frac{1}{2n^{c}}\right )\left(\frac{1}{n} + \frac{1}{n^{c-1}} - \frac{1}{2n^{c}}\right ) < \frac{2}{n} - \frac{1}{n^{c}}$$

$$\frac{1}{n^{c-1}} - \left(\frac{1}{n} - \frac{1}{2n^{c}}\right )\left(\frac{1}{n} + \frac{1}{n^{c-1}} - \frac{1}{2n^{c}}\right ) < 0$$

$$\frac{1}{n^{c-1}} < \left(\frac{1}{n} - \frac{1}{2n^{c}}\right )\left(\frac{1}{n} + \frac{1}{n^{c-1}} - \frac{1}{2n^{c}}\right )$$
$$1 < \left(n^{c-2} - \frac{1}{2n}\right )\left(\frac{1}{n} + \frac{1}{n^{c-1}} - \frac{1}{2n^{c}}\right )$$

$$1 < \left( n^{c-3} - \frac{1}{2n^{2}}\right ) + \left(n^{c-2} - \frac{1}{2n}\right )\left(\frac{1}{n^{c-1}} - \frac{1}{2n^{c}}\right )$$

Note that the first two terms on the RHS are always greater than $1$ for $c > 3$, and the two terms being multiplied are always greater than $0$ for $c \geq 1$. Therefore, this inequality certainly holds. This concludes our proof by contradiction.


\subsection{Proof of Claim~\ref{claim:inner_p3min}}
\label{sec:inner_p3min}
We will prove this by transforming any pair of distributions where $p_{1}(x_{\frac{n}{2} + 1}) \geq \frac{1}{n} - \frac{1}{2n^{c}}$,
such that Player 3 cannot get expected value strictly greater than $1 + \frac{1}{(n-2)n^{c-1}}$, 
into a pair of distributions $p'_1, p'_2$ for the first two players where $p'_{1}(x_{\frac{n}{2}+1}) = \frac{1}{n} - \frac{1}{2n^{c}}$ for the vertex $x_{\frac{n}{2}+1}$ (retaining the same indexing based upon $p_1$), and Player 3 still cannot do better in expectation than this value. 

By construction of the reduction, the condition of Player 3 not being able to get expected value strictly greater than $1 + \frac{1}{(n-2)n^{c-1}}$ by playing vertex moves occurs exactly when the probability that either Player 1 or Player 2 play a vertex move $x_{i}$ is at least $\frac{2}{n} - \frac{1}{n^{c}}$, for all $i$. So, assume that we have a pair of distributions $p_{1}, p_{2}$ of this sort. Then, for all $x_{i}$, $$p_{1}(x_{i}) + p_{2}(x_{i}) - p_{1}(x_{i})p_{2}(x_{i}) \geq \frac{2}{n} - \frac{1}{n^{c}}$$ 

Now, consider any two $x_{j}$, $x_{k}$ s.t. $\frac{2}{n} - \frac{1}{n^{c}} - \epsilon \geq p_{1}(x_{j}) \geq p_{1}(x_{k}) \geq \epsilon$, for $\epsilon>0$. Let us construct a new distribution $p'_{1}$ for Player 1, where everything is the same, except that $p'_{1}(x_{j}) = p_{1}(x_{j}) + \epsilon$, and $p'_{1}(x_{k}) = p_{1}(x_{k}) - \epsilon$. We claim that there will still exist a distribution $p'_{2}$ that Player 2 can play which satisfies $p'_{1}(x_{i}) + p'_{2}(x_{i}) - p'_{1}(x_{i})p'_{2}(x_{i}) \geq \frac{2}{n} - \frac{1}{n^{c}}$ for all $x_i$.

First, note that if $p'_{2}(x_{i}) = p_{2}(x_{i})$ for all $i \neq j,k$, then this condition holds for all these $i$. Then, we only need to ensure that it holds for $j$ and $k$. Player 2 has a total probability of $p_{2}(x_{j}) + p_{2}(x_{k})$ remaining to distribute amongst $p_{2}(x_{j})$ and $p_{2}(x_{k})$. Thus, it is sufficient to prove that there is a valid choice of non-negative values for $p'_{2}(x_{j})$ and $p'_{2}(x_{k})$ such that 
$$p'_{2}(x_{j}) + p'_{2}(x_{k}) \leq p_{2}(x_{j}) + p_{2}(x_{k})$$

We claim that this condition is satisfied by setting the values as follows:

$$p'_{2}(x_{j}) = \frac{\frac{2}{n} - \frac{1}{n^{c}} - p'_{1}(x_{j})}{1 - p'_{1}(x_{j})}$$
$$p'_{2}(x_{k}) = \frac{\frac{2}{n} - \frac{1}{n^{c}} - p'_{1}(x_{k})}{1 - p'_{1}(x_{k})}$$

Which can be rewritten as follows:

$$p'_{2}(x_{j}) = \frac{\frac{2}{n} - \frac{1}{n^{c}} - p_{1}(x_{j}) - \epsilon}{1 - p_{1}(x_{j}) - \epsilon}$$
$$p'_{2}(x_{k}) = \frac{\frac{2}{n} - \frac{1}{n^{c}} - p_{1}(x_{k}) + \epsilon}{1 - p_{1}(x_{k}) + \epsilon}$$

For these values, we have:
$$p'_{2}(x_{j}) + p'_{2}(x_{k}) = \frac{\frac{2}{n} - \frac{1}{n^{c}} - p_{1}(x_{k}) + \epsilon}{1 - p_{1}(x_{k}) + \epsilon} +  \frac{\frac{2}{n} - \frac{1}{n^{c}} - p_{1}(x_{j}) - \epsilon}{1 - p_{1}(x_{j}) - \epsilon}$$

For the original distribution, we use the conditions upon $x_j$ and $x_k$ to get:
$$p_{2}(x_{j}) + p_{2}(x_{k}) \geq \frac{\frac{2}{n} - \frac{1}{n^{c}} - p_{1}(x_{k})}{1 - p_{1}(x_{k})} +  \frac{\frac{2}{n} - \frac{1}{n^{c}} - p_{1}(x_{j})}{1 - p_{1}(x_{j})}$$

So, it is sufficient for us to prove that

$$\frac{\frac{2}{n} - \frac{1}{n^{c}} - p_{1}(x_{k}) +  \epsilon}{1 - p_{1}(x_{k}) + \epsilon} +  \frac{\frac{2}{n} - \frac{1}{n^{c}} - p_{1}(x_{j}) - \epsilon}{1 - p_{1}(x_{j}) - \epsilon}$$ $$ \leq \frac{\frac{2}{n} - \frac{1}{n^{c}} - p_{1}(x_{k})}{1 - p_{1}(x_{k})} +  \frac{\frac{2}{n} - \frac{1}{n^{c}} - p_{1}(x_{j})}{1 - p_{1}(x_{j})} $$

Rearranging terms:
$$ \frac{\frac{2}{n} - \frac{1}{n^{c}} - p_{1}(x_{k}) + \epsilon}{1 - p_{1}(x_{k}) + \epsilon} -  \frac{\frac{2}{n} - \frac{1}{n^{c}} - p_{1}(x_{k})}{1 - p_{1}(x_{k})} $$ $$\leq  +  \frac{\frac{2}{n} - \frac{1}{n^{c}} - p_{1}(x_{j})}{1 - p_{1}(x_{j})} - \frac{\frac{2}{n} - \frac{1}{n^{c}} - p_{1}(x_{j}) - \epsilon}{1 - p_{1}(x_{j}) - \epsilon}$$

$$ \frac{\epsilon(1 - \frac{2}{n} + \frac{1}{n^{c}})}{(1 - p_{1}(x_{k}) + \epsilon)(1 - p_{1}(x_{k}))} \leq \frac{\epsilon(1 - \frac{2}{n} + \frac{1}{n^{c}})}{(1 - p_{1}(x_{j}))(1 - p_{1}(x_{j}) - \epsilon)}$$

Because we know that $1/2 > p_{1}(x_{j}) \geq p_{1}(x_{k}) \geq \epsilon$, all these terms are positive, so we can multiply to get:

$$ (1 - p_{1}(x_{j}) - \epsilon)(1 - p_{1}(x_{j})) \leq (1 - p_{1}(x_{k}) + \epsilon)(1 - p_{1}(x_{k}))$$

Note that the first term on the LHS is no larger than the first term on the RHS, and the second term on the LHS is no larger than the second term on the RHS. Therefore, this statement is always true. This means we have proven that there exists a satisfying distribution $p'_{2}$.

For our given distribution pair $(p_{1}, p_{2})$, we know  that $p_{1}(x_{i}) + p_{2}(x_{i}) - p_{1}(x_{1})p'_{2}(x_{i}) \geq \frac{2}{n} - \frac{1}{n^{c}}$ for all $i$, and $p_{1}(x_{\frac{n}{2} + 1}) \geq \frac{1}{n} - \frac{1}{2n^{c}}$. 
Assume that this is a valid distribution pair that ensures that Player 3 cannot get expected value strictly greater than $1 + \frac{1}{(n-2)n^{c-1}}$. 

We show that we can use the above primitive of moving probability weights repeatedly to get our desired result. Specifically, we show that we can move probability mass from vertex $x_{\frac{n}{2}+1}$ to vertices in the set $V_{1/2} = \{x_1,x_2,\cdots x_{\frac{n}{2}} \}$. To do so, we need to argue that that there are a sufficient number of vertices in $V_{1/2}$ with probability sufficiently below $\delta := \frac{2}{n} - \frac{1}{n^{c}}$ (since this is a necessary condition to be able to move probability mass from a lower probability vertex to a higher probability vertex). Specifically, let $p_1(x_{\frac{n}{2}+1}) = \delta/2 + \gamma$ where $\gamma>0$ (if $\gamma = 0$, we are already done). A simple averaging argument shows that $\gamma < \delta/2$. Assume otherwise, for sake of contradiction. Then, we have $p_1(x_{\frac{n}{2}+1}) \ge \delta$ implying that $p_1(x_i) \ge \delta$ for all $i \in \left[\frac{n}{2} +1 \right ]$. Thus the sum $\sum_i p_1(x_i) \ge \left ( \frac{n}{2} +1 \right) \delta = 1 + \frac{2}{n} - \frac{1}{2n^{c-1}} - \frac{1}{n^c} >1 $, resulting in a contradiction.

Next, we argue that it suffices to show that $\sum_{i=1}^{n/2} p(x_i) \le \frac{n}{2} \delta  - \delta/2$.

We know, by definition of indexing over the $x$'s, that $p_1(x_1) \ge p_1(x_2) \ge p_1(x_3) \cdots p_1(x_n)$. Let $i'$ be the largest index such that $p_1(x_{i'}) \ge \delta$ (with $i' = 0$ if there is no such index). Then, we know that we can move probability mass upto  $\delta - p_1(x_i)$ into each vertex $x_i$ where $i \in \{i'+1,i'+2,\cdots \frac{n}{2}\}$. Let us write $\gamma_i = \delta - p_1(x_i)$ for all $i \in \left [\frac{n}{2}\right]$. Since we assume that $\sum_{i=1}^{n/2} p(x_i) \le \frac{n}{2} \delta  - \delta$, we know that $sum_{i=1} \gamma_i \ge \delta$. This sum can be split into negative and positive parts, which are $\sum_{i=1}^{i'} \gamma_i$ and $\sum_{i=i'
+1}^{\frac{n}{2}} \gamma_i$, implying that $\sum_{i=i'+1}^{\frac{n}{2}} \gamma_i \ge \delta$. We start by moving mass from $x_{\frac{n}{2}+1}$ into $x_{i'+1}$, continue until the probability on $x_{i'+1}$ is $\delta$, and then move onto $x_{i'+2}$ and so on. Since we only need to move $\gamma < \delta/2 $ mass away from $x_{\frac{n}{2}+1}$, we will always be able to complete this procedure. At each step, we maintain the invariant that the probability on $x_{\frac{n}{2}+1}$ is less than or equal to the probability on $x_i$ for $i \in [\frac{n}{2}]$. Thus, to complete the proof, we only need to justify the assumption that $\sum_{i=1}^{n/2} p(x_i) \le \frac{n}{2} \delta  - \delta$.

Since there is no vertex cover of size $\frac{n}{2}$, at least one edge $e$ is left uncovered by the set $V_{1/2}$ (which is a set of $\frac{n}{2}$ vertices). Since, no edge move gets Player 3 a payoff greater than $1 + \frac{1}{(n-2)n^{c-1}}$, there must be at least $\delta$ probability mass of Player 1 picking a vertex that covers edge $e$. Therefore, $\sum_{i=1}^{n/2} p(x_i) \le 1 - \delta $. We prove the stronger result that 

\begin{align*}
    \sum_{i=1}^{n/2} p(x_i) &\le 1 - \delta  
    \le \frac{n}{2} \delta - \delta/2 
\end{align*}
 
To see that the last inequality is true, we substitute the value of $\delta$ on both sides to get :

$$1 - \frac{2}{n} + \frac{1}{n^c} \le 1 - \frac{1}{n} -\frac{1}{2n^{c-1}} + \frac{1}{2n^c}  $$

Rearranging terms gives us : 

$$ \frac{1}{2n^c} + \frac{1}{2n^{c-1}} \le \frac{1}{n} $$
 
Which is true for $c \ge 3$, completing the proof.

Therefore, any valid solution with the constraint given on 
$p_{1}(x_{\frac{n}{2}+ 1})$ implies a valid solution where $p'_{1}(x_{\frac{n}{2}} +1) = \frac{1}{n} - \frac{1}{2n^{c}}$.

\section{Construction of Approximately Optimal $\Gpa$ for Prisoner's Dilemma}
\label{app:pd_construction}

In this section, we will use the ideas in Section~\ref{sec:main_alg} to complete the analysis in Section~\ref{sec:separation}, constructing an approximately optimal Stackelberg strategy for finite-horizon repeated Prisoner's Dilemma. To truly make this construction explicit, we will need to choose a specific $T$. In this case we will pick $T =11$, a value which is not precisely optimal for the coefficients and thus shows off the use of approximation. 

We reproduce the matrix here, scaling down the entries so that payoffs are bounded between $-1$ and $1$:

\[ \begin{bmatrix}
(\frac{3}{5},\frac{3}{5}) & (0,1) \\
(1,0) & (\frac{1}{5}, \frac{1}{5})
\end{bmatrix} \]

As described in section~\ref{sec:main_alg}, constructing the $\Gpa$ involves three main parts: 1) finding the safety value for the Follower, 2) solving an LP to find a distribution, and 3) constructing a $\Gpa$ from that distribution for the particular value of $T$. 

To find the safety value, we can consider the following game matrix which only has the Follower's payoffs, and solve the minimax of the game:
\[ \begin{bmatrix}
\frac{3}{5} & 1 \\
0 & \frac{1}{5}
\end{bmatrix} \]

The solution is $\frac{1}{5}$, and thus we set $V = \frac{1}{5}$. As a sanity check, note that this is the payoff to the Follower when both players defect, which is indeed the highest payoff they can always guarantee themselves. This also gives us the explicit threat that the Leader can use to hit this safety value, which is to play their second move (defect).

To solve the LP shown in section~\ref{sec:LP_UB}, we simply substitute in the values of  $V$ and of the Prisoner's Dilemma payoff matrix (i.e., $M_{1}(1,1) = \frac{3}{5}$, $M_{2}(1,1) = \frac{3}{5}$, and so on). There will be $4$ $\alpha$ variables, as there are $4$ different possible move pairs. This leaves us with the following linear program:  

\begin{flalign*}
    \max \ \ \frac{3}{5}\alpha_{1,1} + \alpha_{2,1} + \frac{1}{5}\alpha_{2,2}  &\\
    \text{ subject to } &\\
     \frac{3}{5}\alpha_{1,1} + \alpha_{1,2} + \frac{1}{5}\alpha_{2,2}  \ge \frac{1}{5} &\\
     \sum_{i,j} \alpha_{i,j} = 1 &\\
    \alpha_{i,j} \ge 0 \qquad \qquad \text{ for } i,j \in [n]^2 &
\end{flalign*}

Solving this linear program gives us the following values:

$\alpha_{1,1} = \frac{1}{3}$
$\alpha_{1,2} = 0$
$\alpha_{2,1} = \frac{2}{3}$
$\alpha_{2,2} = 0$

The value of the LP objective is $\frac{13}{15}$, which also gives us our LP upper bound on the average SE value. 

We are now ready to proceed to the third step, where we construct a $\Gpa$ of size $11$ from this distribution using the tools from section ~\ref{sec:construct_from_LP}. Note that all of the coefficients in the linear program can be written in the form $\frac{p_{i,j}}{3}$, where $p_{i,j} \in \mathbb{N}$. So the cycle length $N$, the LCM of the denominators of $p_{1,1},...p_{2,2}$, is simply $3$. As the LCM is relatively small compared to our choice of $T$, we will try to find our approximate Stackelberg without the sampling approach. So, we simply need to find values for $c$ and $r$ such that $r \leq N$ and
$$T = c \cdot N + r$$
We can achieve this by setting $r = 2$ and $c = 3$. Now, for each action pair, we will calculate the number of times to prescribe playing it in the first $c \cdot N = 9$ rounds:

$$n_{1,1} = \alpha_{1,1} \cdot c \cdot N = \frac{1}{3} \cdot 3 \cdot 3 = 3$$
$$n_{1,2} = \alpha_{1,2} \cdot c \cdot N = 0$$
$$n_{2,1} = \alpha_{2,1} \cdot c \cdot N = \frac{2}{3} \cdot 3 \cdot 3 = 6$$
$$n_{2,2} = \alpha_{2,2} \cdot c \cdot N = 0$$

Now, we will prescribe these action pairs in order of increasing payoff for the Follower. Thus, the prescribed sequence will be $(2,1)$ for the first $6$, rounds, and $(1,1)$ for the next $3$ rounds. Finally, as per the algorithm, the final $r$ rounds will all be the action pair which gets the Follower their highest possible payoff. In this case that is $(1,2)$. So, the final prescribed sequence, which we will call $S^{*}$, is:

$$(2,1),(2,1), (2,1),(2,1),(2,1),(2,1),(1,1),(1,1),(1,1),$$ $$(1,2),(1,2)$$

We can also describe $S^{*}$ it in terms of defection (D) and cooperation (C):
$$(D,C),(D,C), (D,C),(D,C),(D,C),(D,C),(C,C),$$ $$(C,C),(C,C),(C,D),(C,D)$$

For intuition, the first $9$ moves are designed to approximate the payoff of the optimal distribution to give the Follower as close to their best payoff as possible. The final $r = 2$ moves are ``treats'' designed to incentivize the Follower to follow the prescribed strategy up to that point.

To finish the formal construction, the $\Gpa$ will be comprised of $11$ functions $f_{1},...,f_{11}$. In this particular case the threat is deterministic, and therefore no random coins are needed and all outputs are always pure strategies. $f_{1}$ takes no input, and always outputs the first move prescribed to the Leader in $S^{*}$. The remaining functions $f_{i}$ all operate as follows: check the history of play of length $i-1$ against the first $i-1$ move pairs in $S^{*}$. If it matches perfectly, play the $i$th move prescribed to the Leader in $S^{*}$. If it does not match perfectly, play $D$ (the threat). 

By the correctness proofs in ~\ref{sec:construct_from_LP}, a best-responding Follower will obey this prescribed sequence. Therefore we can compute the Leader's average payoff to be $$\frac{1 \cdot 6 + \frac{3}{5} \cdot 3 + \frac{1}{5} \cdot 2}{11} = \frac{8}{11}$$ 

Recall that the LP upper bound is $\frac{13}{15}$. So, the approximation gap in this instance is 

$$\frac{13}{15} - \frac{8}{11} = \frac{23}{165}$$

We have claimed that this approach should guarantee a $\frac{2r}{T}$-additive approximation to the optimal, which we can now check. We have $r = 2$, $T = 11$. So, we hope to guarantee a $\frac{4}{11}$-approximation in this case. Indeed, $\frac{23}{165} < \frac{4}{11}$, and thus our construction indeed gets the correct approximation guarantee.

\section{No-Regret Leader Strategies}
\label{app:nr}
The approximately optimal Stackelberg GPA that we have described in this paper is appealing in that it is easy to compute and easy for the Follower to reason about. However, the algorithm which computes this GPA does require the entire game matrix as input. One might wonder whether no-regret algorithms, which require limited information about the payoff matrix, are also good approximations to Stackelberg $\Gpa$s. 




In section~\ref{sec:zero_sum}, we prove that such an approach will in fact work for two player zero-sum games. We first show how the distinction between single-shot and repeated games, as well as the distinction between Nash equilibria and SE, collapses in this setting, and use these results to prove that any no-regret algorithm is an approximately optimal Stackelberg GPA in a two player zero-sum game.

However, for general-sum games, it is not sufficient to construct Stackelberg GPAs from learning algorithms. In section~\ref{sec:gen_sum}, we show that there exist general-sum games where being an approximately optimal Stackelberg GPA is incompatible with having the no-regret guarantee. This result motivates the need for our GPA-producing algorithm to require full information upfront.


We remain in the same setting of finite-horizon repeated two-player games that we have been discussing throughout this paper. However, we must introduce some additional notation to describe the mechanics of the no-regret algorithms. 

\begin{definition}[Regret]
Let $x_1,x_2 \cdots x_T \in \Delta^n$ be the randomized strategies played by some player in $t$ rounds of play. The expected payoff of this algorithm is therefore $\sum_{t=1}^T x_t^\intercal p_t$. Consider the payoff of playing a single strategy $i$ in every round against the same payoff vectors generated by the environment --- we denote the probability distribution on support $[n]$ that puts the entire mass on $i$ as $e_i$. This payoff is $\sum_{t=1}^T e_i^\intercal p_t$. We say that the algorithm has regret $r(T,n)$ if $\sum_{t=1}^T e_i^\intercal p_t - \sum_{t=1}^T x_t^\intercal p_t \le r(T,n)$ for every possible environment. We also define the notion of the regret of an algorithm for a specific class of environments, when the above condition is only guaranteed to hold for those specific environments.
\end{definition}

\begin{definition}[No-Regret]
We say that an algorithm is No-Regret if $\frac{r(T,n)}{T} \rightarrow 0$ as $T \rightarrow \infty$.
\end{definition}

No-Regret algorithms are known to exist (See~\cite{cesa2006prediction} for a survey). The optimal algorithms for minimizing regret in arbitrary online environments have $r(T,n) = \Tilde{O}(\sqrt{T})$.

\begin{definition}
[Value of a Game]
In a two player zero-sum game, the value $v$ of a game is the expected payoff that the maximizing player will get in all of their Nash Equilibria. Equivalently, $v$ is the maximin / minimax value of the game (the equivalence of these two values is the Von-Neumann Minimax Theorem).
\end{definition}

\subsection{Stackelberg No Regret in Zero-Sum Games}
\label{sec:zero_sum}
In zero-sum games, the distinctions between NE and SE no longer hold. Similarly, the distinctions between equilibria in one-shot games and equilibria in repeated games no longer hold. The simplicity of the zero-sum games setting allows for much simpler Stackelberg algorithms, such as no-regret algorithms. In the proofs below, we will assume that Player 1 is the maximization player and Player 2 is the minimization player. Furthermore, when relevant, Player 1 is the Leader. We will utilize the following well-known fact:
\begin{lemma}
\label{lem:nr_is_stack}
In single-shot two player zero-sum games, the following three values are equivalent: the Nash equilibrium value for Player p, the SE value for Player p if they are the Leader, and the SE value for Player p if they are the Follower. If player p is the min player, then this value is also equal to the minimax of the game. If player p is the max player, this value is also equal to the maximin of the game.
\end{lemma}

Using this, we will prove the following lemma:

\begin{lemma}
In zero-sum repeated games, the average Stackelberg GPA value is the same as the single-shot Stackelberg value.
\end{lemma}
\begin{proof}
Assume for contradiction that there exists a Stackelberg GPA which gets the maximization player some value strictly greater than $v$ on average. The minimization player/Follower could always ensure that the value is $v$ on average by playing their half of a Nash Equilibrium. Thus the Follower is not best responding, which derives a contradiction. Therefore the Stackelberg GPA value is no higher than $v$. In addition, there exists a GPA which will get the Follower value $v$: simply playing the single-shot Stackelberg strategy (or, equivalently, any Nash Equilibrium). Thus, the Stackelberg GPA value must be exactly $v$, the same as the single-shot Stackelberg value.    
\end{proof}

\begin{corollary}
\label{cor:lots_of_equivalences}
In zero-sum, single-shot, two-player games, the following values are all equivalent for player p: the maximin value of the game, the single-shot Nash equilibrium value for p, the single-shot SE value for p (both as Leader and Follower), and the repeated SE value for p (both as Leader and Follower). 
\end{corollary}

We will now show the equivalence between Stackelberg and Nash in repeated zero-sum games, which holds even in approximation. Recall that in an $\epsilon$-approximate Nash equilibrium, both players may be $\epsilon$-close to best responding. By contrast, in an $\epsilon$-approximate SE, the Follower must be exactly best-responding, while the Leader may be getting $\epsilon$-close to their optimal SE.

\begin{lemma}
\label{lem:nr_is_stack_repeated}
In zero-sum repeated games, every GPA which is part of some $\epsilon$-approximate Nash Equilibrium is a $2\epsilon$-approximate Stackelberg GPA. In addition, every GPA which is a $\epsilon$-approximate Stackelberg GPA is part of some $\epsilon$-approximate Nash Equilibrium. 
\end{lemma}
\begin{proof}
First, consider any GPA $g$ which is the maximization player's half of an $\epsilon$-approximate Nash Equilibrium. Note that regardless of what the minimization player is doing, the maximizing player could simply play their half of a one-shot Nash equilibrium every round and guarantee themselves $T \cdot v$ in expectation. Therefore, in order to have an $\epsilon$-approximate Nash on average, it must be that the maximization player is getting expected value at least $T \cdot (v - \epsilon)$. Next, note that the minimizing player is within $\epsilon$ of best-responding on average to $g$. Thus, if the minimizing player did best-respond to $g$, the most that they could decrease the payoff by would be $\epsilon \cdot T$. Therefore, if the Leader in a Stackelberg game played $g$, a perfectly best-responding minimizing player would make the payoff for Player 1 be no less than $T \cdot (v - 2\epsilon)$. By lemma 12, the best that the Leader could possibly do is get $v$. Therefore, $g$ is a $2 \epsilon$-approximate Stackelberg GPA. 

Next, consider any GPA $g$ which is an $\epsilon$-approximate Stackelberg GPA. Let the opponent's best response be some GPA $f$. Recall that there is a GPA which gets Player 1 value $T \cdot v$. So, the definition of approximate SE, when Player 2 plays $f$ against $g$, Player 1 gets at least $T \cdot (v - \epsilon)$ in expectation. By Corollary~\ref{cor:lots_of_equivalences}, Player 1 can never get payoff higher than $T \cdot v$. Therefore, $g$ is a $\epsilon$-approximate best response to $f$ on average, and $f$ is an exact best response to $g$. Thus they form an $\epsilon$-approximate Nash equilibrium.
\end{proof}

We now prove the main result of this subsection, and show that no-regret algorithms are approximate Stackelberg $\Gpa$s.

\begin{lemma}
\label{lemma:zs_nr_st}
In two player zero-sum games, any GPA with regret $r(T,n)$ is a $\left(\frac{r(T,n)}{T}\right)$-approximate Stackelberg GPA.
\end{lemma}
\begin{proof}
 We know that the Follower, the minimization player, can always guarantee themselves a value of at most $v \cdot T$ by playing their half of any Nash Equilibrium at each round. Thus, regardless of what Player 1 commits to, they can do no better than an expected payoff of $v \cdot T$. Now, consider the transcript generated by Player 1 using the no-regret algorithm and Player 2 best-responding. The best action in hindsight for Player 1 will definitely get them at least the game value $v$ (on average over rounds) in expectation, since it can be viewed as best responding to the historical average transcript of Player 2. Therefore, the no-regret property guarantees that the expected payoff of Player 1 is t least $v \cdot T - r(T,n)$. 
 Therefore, the average expected payoff difference between this no regret GPA and the Stackleberg GPA is at most $\frac{r(T,n)}{T}$.
\end{proof}

\subsection{Failure of No-Regret Leader GPAs in General-Sum Games}
\label{sec:gen_sum}
While committing to a no-regret algorithm is approximately optimal in two player zero-sum games, this result does not extend to the general-sum case. In fact, the contrast is quite sharp; \emph{any} no-regret algorithm will be an approximate Stackleberg GPA to \emph{any} two player zero-sum game. But there exist two player general-sum games such that there is \emph{no} no-regret algorithm which is an approximate Stackeberg GPA. 


\begin{lemma} There exist two-player, general-sum games such that every algorithm which is an $\alpha$-approximate Stackelberg strategy for $\alpha \leq c$ will will incur average regret of at least $\frac{c}{8}$. Here, $c$ is some constant.
\end{lemma}

Readers may note that the proof holds for the stronger class of no-swap regret algorithms as well.

\begin{proof}
Consider the following game matrix M, where Player 1 (the Leader) plays the rows and Player 2 (the Follower) plays the columns:
\[ \begin{bmatrix}
(\frac{1}{4},1) & (\frac{3}{4}, 0) \\
(0,0) & (\frac{1}{2}, 1)
\end{bmatrix} \]

We show that in this game, \emph{any} $o(1)$-approximate Stackelberg strategy for Player 1 has $\Omega(T)$ regret.

\begin{claim}
There exists some value $T$ such that for any game of length $T' \geq T$, there is no GPA which is both a $\frac{1}{8}$-approximate Stackelberg GPA and gets the Leader total regret below $\frac{1}{64} \cdot T$.
\end{claim}

Before diving into the proof, let us try to provide some intuition for why this might be the case. Consider a simple GPA $g$ for the Leader where they commit to always playing the first row. The optimal response for the Follower will be to always play the first column. Then the payoff for the Leader will be $\frac{1}{4}$. $g$ has regret of $0$, because the first row is a dominant strategy. Now, consider GPA $g'$, where the Leader always plays the second row. The Follower will best respond by always playing the second column. So the payoff for the Leader will be $\frac{1}{2}$. However, note that against the static transcript of play, the Leader now has $\Omega(T)$ regret compared to switching to row 1.

Consequently, $g'$ has much better payoff than $g$, but also much higher regret. There seems to be an inherent tradeoff for the Leader between reducing regret by usually playing row 1 and increasing payoff by usually playing row 2. We formalize this below and show that, in fact, the group of algorithms which allow for low regret (like $g$) are fully disjoint from the group of algorithms which are near-best Stackelberg GPAs (like $g'$).

Specifically, we will consider the Leader's strategy $g^{*}$, for which there is always a well-defined best response $f^{*}$. These GPAs can be simulated against each other to produce a probability distribution over outcomes. We use $p_{r_i,c_j}$ to denote the probability of the move pair of row $i$ and column $j$, averaged over all rounds and over the entire probability distribution. We similarly denote by $p_{r_i}$ the probability of the Leader playing row $i$, regardless of the Follower's move. We will prove that if $p_{r_2} \geq \frac{1}{16}$, they will have $\Omega(T)$ regret. Then, we will show that if $p_{r_2} \leq \frac{1}{16}$, the strategy is at least $\frac{1}{8}$-far from a SE. Thus, every strategy is either a constant far from a SE, or induces constant regret. \\

Part 1: If $p_{r_2} \geq \frac{1}{16}$, the historical transcript will have $\Omega(T)$ regret in expectation.

Proof:
The expected payoff of the Leader in any given run of the game is 

$$E_{s \sim D}[u(s)] = \frac{p_{(r1,c1)} }{4} + \frac{3p_{(r1,c2)} }{4} + \frac{p_{(r2,c2)}}{2}$$

Consider instead playing row $1$ against any given historical transcript from a run of the game. Call this strategy $s'$. Then, their expected payoff would be
$$E_{s \sim D}[u(s')] = \frac{p_{(r1,c1)} + p_{(r2,c1)}}{4} + \frac{3(p_{(r1,c2)} + p_{(r2,c2)})}{4}$$
$$= \frac{p_{(r1,c1)}}{4} + \frac{3(p_{(r1,c2)})}{4}+ \frac{p_{(r2,c2)}}{2} + \frac{p_{(r2,c2)}+ p_{(r2,c1)}}{4}$$
$$= E_{s \sim D}[u(s)] + \frac{p_{(r2,c2)}+ p_{(r2,c1)}}{4}$$

From our assumption, we know that $p_{r2,c1} + p_{r2,c2} \geq \frac{1}{16}$, so we have:
$$E_{s \sim D}[u(s')] \geq  E_{s \sim D}[u(s)] + \frac{1}{64}$$
Thus, the Leader's regret is at least $\frac{1}{64}$, as the strategy of always playing row $1$ is exactly what gets the payoff improvement. \\

Part 2: If $p_{r_2} \leq \frac{1}{16}$ and the Leader's regret is $\leq \frac{1}{64}$, then $g^{*}$ $\frac{1}{8}$-far from a SE. The intuition here is that if the Leader is somehow able to get close to the optimal value by playing almost entirely the first row, it must be because the Follower is often playing the second column. But if Player 2 is often playing the second column while Player 1 plays the first row, then the Follower is not best-responding, deriving a contradiction. We go over this idea more carefully below. \\
Note that there exists a GPA for the Leader that garners an average payoff of $\frac{1}{2}$: simply committing to always play row $2$. Then the Follower must always play column 2 in order to exactly best-respond. Thus, the Leader will get expected value of $\frac{1}{2}$. 
\\
Thus, this is a lower bound on the payoff of the Stackelberg GPA. In order to get within $\frac{1}{8}$ of a best response, then, a strategy must have expected payoff of at least $\frac{1}{2} - \frac{1}{8} = \frac{3}{8}$. \\
Assume for contradiction that there exists some $g^{*}$ such that the expected payoff is at least $\frac{3}{8}$ and the second row is on average played less than $\frac{1}{16}$ of the time.  \\
We can examine the history of play and look at the frequency of each move combination: let us call them $f_{11}$, $f_{12}$,$f_{21}$ and $f_{22}$. We can write the expected payoff for the Leader as a linear combination of these. Since we know the payoff is at least $\frac{3}{8}$, we can write:
$$\frac{3}{8} \leq \frac{1}{4}f_{11} + \frac{3}{4}f_{12} + 0 \cdot f_{21} + \frac{1}{2}f_{22}$$
$$ \frac{3}{8} \leq \frac{1}{4}f_{11} + \frac{3}{4}f_{12} + \frac{1}{2}f_{22}$$

Using the fact that row $2$ is played less than $\frac{1}{16}$ of the time, we know that $f_{22} \leq \frac{1}{16}$. So we get:
$$ \frac{3}{8} \leq \frac{1}{4}f_{11} + \frac{3}{4}f_{12} + \frac{1}{32}$$

Finally, let's use the fact that $f_{11} + f_{12} \leq 1$:
$$ \frac{3}{8} \leq \frac{1}{4} + \frac{1}{2}f_{12} + \frac{1}{32}$$
$$ \frac{3}{32} \leq \frac{1}{2}f_{12}$$
$$ \frac{3}{16} \leq f_{12}$$

The payoff of the Follower is $$f_{11} + f_{22}$$
We can upper bound this payoff in our current setting by upper bounding these values, using the fact that $f_{12} \geq \frac{3}{16} $. This means that $f_{11} \leq \frac{13}{16}$. Also, we have that $ f_{21} + f_{22} \leq \frac{1}{16}$. Thus means that $f_{22} \leq \frac{1}{16}$. So, the total Follower payoff is at most $\frac{14}{16}$. \\
Now, consider the following pure strategy for the Follower: always play column $1$. From the proof in case 1, as the Leader is playing an algorithm with regret at most $\frac{1}{64}$, $g^{*}$ must play row $2$ at most $\frac{1}{16}$ of the time in expectation over induced transcripts against this Follower strategy. This means that the expected payoff for the Follower for this strategy is at least $\frac{15}{16}$. This is a $\frac{1}{16}$-improvement in expectation over the payoff of $f^{*}$, which means that $f^{*}$ is not a best response. Therefore we have arrived at a contradiction. This means that in this case, the Leader algorithm is $\frac{1}{8}$-far from being a best response. 
\end{proof}

\end{document}